\documentclass[preprintnumbers,preprint,aps,prd,amsmath]{revtex4}

\usepackage{bm}
\usepackage{amsmath}
\usepackage[dvips]{color,graphicx}
\usepackage{epsfig}
\usepackage[figuresright]{rotating}
\usepackage{subfigure}
\usepackage{hhline,multirow}  % for nicer tables
\usepackage{dcolumn}          % Align table columns on decimal point
\usepackage{color}
\usepackage[colorlinks=true,linkcolor=blue]{hyperref}

\newcommand{\eps}{\varepsilon}
\newcommand{\pzhatsq}{{\widehat{p}}_z^2}

\begin{document}

\preprint{JLAB-THY-13-1770}

\title{Comparative study of nuclear effects in polarized \\
	electron scattering from $^3$He}
\author{J. J. Ethier$^{1,2}$, W. Melnitchouk$^2$}
\address{$^1$\mbox{Physics Department, Stetson University,
		DeLand, Florida 32723}	\\
	 $^2$\mbox{Jefferson Lab, 12000 Jefferson Avenue,
		Newport News, Virginia 23606}}
\date{\today\\}

\begin{abstract}
We present a detailed analysis of nuclear effects in inclusive
electron scattering from polarized $^3$He nuclei for polarization
asymmetries, structure functions and their moments, both in the
nucleon resonance and deep-inelastic regions.
We compare the results of calculations within the weak binding
approximation at finite $Q^2$ with the effective polarization
{\it ansatz} often used in experimental data analyses, and explore
the impact of $\Delta$ components in the nuclear wave function
and nucleon off-shell corrections on extractions of the free
neutron structure.
Using the same framework we also make predictions for the $Q^2$
dependence of quasielastic scattering from polarized $^3$He,
data on which can be used to constrain the spin-dependent nuclear
smearing functions in $^3$He.
\end{abstract}

\maketitle

%%%%%%%%%%%%%%%%%%%%%%%%%%%%%%%%%%%%%%%%%%%%%%%%%%%%%%%%%%%%%%%%%%%%%%%%%
\section{Introduction}

Reliable extraction of information on the spin structure of the
neutron is vital for our understanding of the flavor and spin
decomposition of the nucleon in terms of its quark and gluon
constituents.  When combined with the more copious measurements
of the proton structure, the neutron data allow the individual
$u$ and $d$ flavor contributions to be determined.  The absence
of free neutron targets, however, means that polarized nuclei such
as deuterium, $^3$He or $^7$Li must be used as effective polarized
neutron targets.
The perennial problem of nuclear corrections must therefore be
seriously addressed if one is obtain neutron structure information
with sufficient accuracy.  This is especially pertinent for new
generations of polarized deep-inelastic scattering (DIS) experiments
that aim to measure spin structure functions and their moments with
unprecedented precision \cite{Che11, Aid12}.

The study of nuclear corrections to spin structure functions in the
DIS and quasi-elastic (QE) regions has some history already, with
the first quantitative calculations dating back to the 1980s
\cite{FS83, Blank84, Wol89}.  Subsequent work by Ciofi~degli~Atti
{\it et al.} \cite{Ciofi92, Ciofi93, Ciofi95, Ciofi97} and other
groups \cite{Kaptari90, SS, KPSV, Bissey01, Boros01, BGST02} over the
past two decades has made important inroads into our understanding of
nuclear effects in scattering from polarized $^3$He nuclei, and the
extraction of the spin structure of the free neutron.  The technology 
developed for inclusive scattering has more recently been extended to
other observables, such as the $^3$He generalized parton distributions
\cite{ScopGPD} and transversity in semi-inclusive DIS \cite{ScopSIDIS}.

While most of the traditional approaches have been based on a
nonrelativistic treatment of the dynamics, some calculations of
polarized $^3$He structure functions have attempted to incorporate
relativistic and off-shell effects \cite{MPT, PMT, SSlc}, although
these are generally difficult to constrain unambiguously.
A systematic approach for expanding the nucleon propagator in the
nuclear medium was developed \cite{KPW, KMPW} in the weak binding
approximation (WBA), in which the usual convolution formulas for
nuclear structure functions can be derived to order $\bm{p}^2/M^2$,
where $\bm{p}$ is the nucleon three-momentum and $M$ is its mass,
with identifiable higher order corrections.
The WBA method was applied to polarized deuteron \cite{KMd} and
$^3$He nuclei \cite{KMhe}, where the nuclear corrections were
estimated in both the DIS and nucleon resonance regions.

The standard nuclear structure function analyses have for the most
part been formulated within the plane wave impulse approximation,
in which scattering is assumed to take place incoherently from
individual nucleons within the nucleus, with the closure
approximation used to sum over the hadrons in the final state.
Beyond the impulse approximation, nuclear shadowing corrections
in polarized $^3$He have been shown to arise from multiple
scattering of the lepton from two or more nucleons in the $^3$He
nucleus \cite{Frank96, Guzey99}.
In addition, contributions to the spin-dependent $^3$He structure
function from non-nucleonic degrees of freedom in the nucleus,
such as the $\Delta(1232)$ isobar, have been argued \cite{Boros01,
BGST02} to account for the $\approx 4\%$ difference between the
value of the isovector axial charge $g_A$ in the free nucleon
(measured in neutron $\beta$ decay) and that in the $A=3$ nuclei
(from tritium $\beta$ decay) \cite{Budick91}.

With recent polarized $^3$He experiments at Jefferson Lab, as well
as those planned for the upcoming 12~GeV energy upgrade, attaining
ever greater precision, the need exists for increasingly accurate
theoretical descriptions of the nuclear corrections to spin-dependent
structure functions and their moments.
For example, the $d_2$ moment of the $^3$He structure functions has
recently been measured in the E06-014 experiment \cite{E06-014}
at Jefferson Lab Hall~A, which can in principle reveal certain
higher-twist matrix elements of the neutron --- provided the
nuclear corrections can be accounted for.
Furthermore, extraction of the neutron polarization asymmetry
$A_1^n$ from data on the $A_1^{^3{\rm He}}$ asymmetry requires
the simultaneous determination of nuclear corrections to the
spin-dependent $g_1$ and $g_2$ structure functions, as well as
the unpolarized $F_1$ and $F_2$ structure functions of $^3$He,
which has not been systematically considered in previous work.

In this paper we revisit the problem of nuclear effects in inclusive
scattering of polarized leptons from polarized $^3$He nuclei in the
DIS, nucleon resonance and QE regions, focusing in particular on
kinematics at intermediate and large values of Bjorken $x$.
We provide a critical comparison of various approaches and
approximations to computing the nuclear corrections, with a view of
obtaining a more reliable estimate of the uncertainty on the nuclear
effects to be used in extractions of the free neutron structure.
We work within the framework of the WBA to compute the polarized
nucleon light-cone momentum distributions in $^3$He (commonly referred
to as ``smearing functions''), and compare the full results at finite
four-momentum transfer squared $Q^2$ with those often used in the
large-$Q^2$ approximation, as well as with the effective polarization
{\it ansatz} typically adopted in experimental data analyses.
In addition to the standard nuclear smearing corrections which
incorporate Fermi motion and binding effects, we discuss the effects
of non-nucleonic constituents of the nucleus such as the $\Delta$
resonance, and the possible off-shell modification of the nucleon
structure functions in the nuclear medium.
The effects of the various nuclear corrections are considered for
both the $g_1$ and $g_2$ structure functions and their moments,
as well as for the $A_1$ and $A_2$ polarization asymmetries,
which requires correcting also the unpolarized $^3$He structure
functions.
As a possible test of the reliability of the nuclear corrections,
we evaluate the QE contributions to the spin structure functions,
which can be compared with future precision data from dedicated
$^3$He experiments in the QE region.

This paper is organized as follows.
In Sec.~\ref{sec:def} we summarize the basic formulas for the
inclusive cross sections, structure functions and polarization
asymmetries relevant for the analysis.  The formalism used
for computing the polarized $^3$He structure functions within the
WBA is outlined in Sec.~\ref{sec:nuclear}, where we discuss the
full results for the nuclear smearing functions at finite values of
$Q^2$, as well as those in the Bjorken limit and in the zero-width
approximation.  The latter leads to the effective polarization
approximation, which is often used in analyses of
$^3$He data.  Numerical results for structure functions, asymmetries
and moments are presented in Sec.~\ref{sec:results}, where we study
the dependence on the nuclear wave function, and test the efficacy of
the various approximations.  The possible impact of effects beyond
the impulse approximation, namely, from $\Delta$ degrees of
freedom and nucleon off-shell corrections, is also examined.
Finally, in Sec.~\ref{sec:conc} we summarize our findings
and discuss possible further applications of this work.

%%%%%%%%%%%%%%%%%%%%%%%%%%%%%%%%%%%%%%%%%%%%%%%%%%%%%%%%%%%%%%%%%%%%%%%%%
\section{Cross sections and asymmetries}
\label{sec:def}

To begin our discussion we first summarize the main formulas for
cross sections and polarization asymmetries in terms of the
spin-dependent $g_1$ and $g_2$ structure functions.
The structure functions can be extracted from measurements of
longitudinally polarized leptons scattered from a target that is
polarized either longitudinally or transversely relative to the
electron beam.
For longitudinal beam and target polarization, the difference between
the spin-aligned and spin-antialigned cross sections (with the arrows
$\uparrow$ and $\Uparrow$ denoting the electron and nucleon spin
orientations, respectively) is given in the target rest frame by
\begin{eqnarray}
\label{eq:sig_long}
{ d^2\sigma^{\uparrow\Downarrow} \over d\Omega dE' }  
- { d^2\sigma^{\uparrow\Uparrow} \over d\Omega dE' }
&=& { \sigma_{\rm Mott} \over M \nu }
  4 \tan^2{\theta \over 2}
  \Big[ (E+E'\cos\theta) g_1(x,Q^2) - 2 M x g_2(x,Q^2)
  \Big],
\end{eqnarray}
where $E$ and $E'$ are the incident and scattered electron energies,
$\nu = E-E'$ is the energy transfer and $\theta$ is the electron
scattering angle.  The Bjorken scaling variable is defined as
$x = Q^2/2M\nu$, and
$\sigma_{\rm Mott} = (4\alpha^2 E'^2/Q^4) \cos^2(\theta/2)$
is the Mott cross section for scattering from a point particle.
The $g_2$ structure function can be extracted if one in addition
measures the cross section for a nucleon polarized in a direction
transverse to the beam polarization,
\begin{eqnarray}
\label{eq:sig_transv}
{ d^2\sigma^{\uparrow\Rightarrow} \over d\Omega dE' }
- { d^2\sigma^{\uparrow\Leftarrow} \over d\Omega dE' }
&=& { \sigma_{\rm Mott} \over M \nu }
  4 \tan^2{\theta \over 2}\, E' \sin\theta
  \Big[ g_1(x,Q^2) + { 2 E \over \nu } g_2(x,Q^2) \Big].
\end{eqnarray}

In practice, it is often easier to measure polarization asymmetries,
or ratios of spin-dependent to spin-averaged cross sections.
The ratios of the cross section differences in Eqs.~(\ref{eq:sig_long})
and (\ref{eq:sig_transv}) to the sums of the cross sections define the
longitudinal $A_{\|}$ and transverse $A_{\perp}$ polarization
asymmetries, respectively.
%
% \begin{eqnarray}
% \label{eq:Apar} 
% A_{\|}
% &=& { \sigma^{\uparrow\Downarrow} - \sigma^{\uparrow\Uparrow}
% \over \sigma^{\uparrow\Downarrow} + \sigma^{\uparrow\Uparrow} }\ ,   \\
%
% \label{eq:Aperp}
% A_{\perp}       
% &=& { \sigma^{\uparrow\Rightarrow} - \sigma^{\uparrow\Leftarrow}
% \over \sigma^{\uparrow\Rightarrow} + \sigma^{\uparrow\Leftarrow} }\ ,
% \end{eqnarray}
%
% where for shorthand we denote              
% $\sigma^{\uparrow\Downarrow}
% \equiv d^2\sigma^{\uparrow\Downarrow}/d\Omega dE'$, {\em etc.}
%
The $g_1$ and $g_2$ structure functions can then be extracted from
these asymmetries according to
\begin{subequations}
\label{eq:g1}
\begin{eqnarray}
g_1(x,Q^2)
&=& F_1(x,Q^2)\, { 1 \over d'}
  \left[ A_{\|} + \tan{\theta \over 2} A_{\perp} \right],       \\
\label{eq:g2} 
g_2(x,Q^2)
&=& F_1(x,Q^2)\, { y_e \over 2d' }
  \left[ { E + E' \cos\theta \over E' \sin\theta } A_{\perp} - A_{\|}   
  \right],
\end{eqnarray}
\end{subequations}
where the kinematical factor
$d' = (1 - \epsilon)(2 - y_e) / [y_e (1 + \epsilon R(x,Q^2))]$,
$\epsilon$ is the ratio of longitudinal to transverse virtual
photon polarizations, and $y_e = \nu/E$ is the fractional energy
loss of the incident electron.
The ratio $R$ is defined in terms of the spin-averaged longitudinal
and transverse structure functions,
\begin{equation}
\label{eq:Rfl}
R(x,Q^2) = { F_L(x,Q^2) \over 2x F_1(x,Q^2) },
\end{equation}
where
\begin{equation}
\label{eq:flsigl}
F_L(x,Q^2) = \gamma^2 F_2(x,Q^2) - 2x F_1(x,Q^2),
\end{equation}
with $\gamma^2 = \bm{q}^2/\nu^2 = 1 + 4 M^2 x^2/Q^2$ and
$\bm{q}^2 = \nu^2 + Q^2$.
Note that the $F_1$ structure function is related only to the transverse
virtual photon coupling, while $F_2$ is a combination of both transverse
and longitudinal couplings.

One can also define virtual photon absorption asymmetries $A_1$ and
$A_2$ in terms of the measured asymmetries,
\begin{subequations}
\label{eq:Apar12}
\begin{eqnarray}
A_{\|} &=& D (A_1 + \eta A_2),        \\
\label{eq:Aperp12}
A_{\perp} &=& d (A_2 - \zeta A_1),
\end{eqnarray}
\end{subequations}
where $D = (1 - E' \epsilon/E)/(1 + \epsilon R(x,Q^2))$
is the photon depolarization factor,
and the other kinematic factors are given by
$\eta = \epsilon \sqrt{Q^2}/(E - E' \epsilon)$,
$d = D\sqrt{2 \epsilon /(1 + \epsilon)}$,
and $\zeta = \eta(1 + \epsilon) / 2\epsilon$.
%
% A_1 = { d A_\parallel - D \eta A_\perp \over
%         d D (1 + \zeta \eta) }
%
% A_2 = { d \zeta A_\parallel + D A_\perp \over
%         d D (1 + \zeta \eta) }
%
The $A_1$ and $A_2$ asymmetries can also be directly expressed in terms
of the $g_1$, $g_2$ and $F_1$ structure functions,
\begin{subequations}
\label{eq:asym}   
\begin{eqnarray}
A_1(x,Q^2) &=& {1 \over F_1(x,Q^2)}
\left[ g_1(x,Q^2) - (\gamma^2-1) g_2(x,Q^2) \right],
\label{eq:A1}						\\
A_2(x,Q^2) &=& {\sqrt{\gamma^2-1} \over F_1(x,Q^2)}
\left[ g_1(x,Q^2) + g_2(x,Q^2) \right].
\label{eq:A2}
\end{eqnarray}
\end{subequations}
At small values of $x^2/Q^2$, one then finds $A_1 \approx g_1/F_1$.
In the same limit, the $A_2$ asymmetry also vanishes: $A_2 \to 0$
for $\gamma \to 1$.
If the $Q^2$ dependence of the polarized and unpolarized structure
functions is similar, the polarization asymmetry $A_1$ will be
weakly dependent on $Q^2$.
Furthermore, positivity constrains lead to bounds on the magnitude
of the virtual photon asymmetries,
\begin{eqnarray}
|A_1| \le 1\ ,\ \ \ \ \ |A_2| \le \sqrt{R}\ .
\end{eqnarray}

For QCD analysis it is usually convenient to work in terms of
(Cornwall-Norton) moments of the $g_1$ and $g_2$ structure functions;
the $n$-th moments are defined as
\begin{eqnarray}  
\label{eq:moment}
\Gamma_{1,2}^{(n)}(Q^2)
&=& \int_0^1 dx\ x^{n-1} g_{1,2}(x,Q^2).
\end{eqnarray}
Note that since the moments integrate the structure functions up to
$x=1$, they formally include the elastic scattering contributions.
The elastic contributions to $g_1$ and $g_2$ can be written in terms
of the Sachs electric and magnetic form factors as
\begin{subequations}
\label{eq:QE}
\begin{eqnarray}
g_1^{(\rm el)}(x,Q^2)
&=& {1 \over 2(1+\tau)}
    G_M(Q^2) \left( G_E(Q^2) + \tau\, G_M(Q^2) \right)
    \delta(x-1),                        \\
g_2^{(\rm el)}(x,Q^2)
&=& {\tau \over 2(1+\tau)}
    G_M(Q^2) \left( G_E(Q^2) - G_M(Q^2) \right)
    \delta(x-1),
\end{eqnarray}
\end{subequations}
where $\tau = Q^2/4M^2$.
Of particular interest to the study of the nucleon's nonperturbative
structure is the $d_2$ moment,
\begin{eqnarray}
\label{eq:d2}
d_2(Q^2)
&=& \int_0^1 dx\, x^2\, \Big( 2 g_1(x,Q^2) + 3 g_2(x,Q^2) \Big),
\end{eqnarray}  
which is defined so as to expose the twist-3 part of the $g_2$ structure
function.  The $g_2$ structure function is unique amongst the nucleon's  
structure functions in that its higher twist contributions are not
suppressed by powers of $1/Q^2$, but enter at the same order as the
twist-2 component.  The latter is given by the Wandzura-Wilczek
relation \cite{WW},
\begin{eqnarray}
g_2^{\rm WW}(x,Q^2)
= - g_1(x,Q^2) + \int_x^1 \frac{dz}{z}\, g_1(z,Q^2),
\label{eq:WW}
\end{eqnarray}
where $g_1$ here includes only twist-2 contributions.
In general, the total $d_2$ moment can be written in terms
of the twist-2 (WW) and higher twist contributions,
$d_2 = d_2^{\rm WW} + \overline{d}_2$.
Using Eq.~(\ref{eq:WW}) one can verify that the twist-2 part
of $d_2$ vanishes, $d_2^{\rm WW} = 0$, so that measurement of
$d_2$ cleanly reveals the higher twist component $\overline{d}_2$.
Obviously, for the WW component the lowest moment of $g_2$
vanishes identically, $\Gamma_2^{(1) {\rm WW}} = 0$ \cite{BC70}.

%%%%%%%%%%%%%%%%%%%%%%%%%%%%%%%%%%%%%%%%%%%%%%%%%%%%%%%%%%%%%%%%%%%%%%%%%
\section{Nuclear structure functions}
\label{sec:nuclear}

In this section we present the formalism for computing the
spin-dependent structure functions of $^3$He, and discuss their
relation to the spin structure functions of the proton and neutron.
Within the weak binding approximation, the nuclear and nucleon
structure functions can be related by convolutions involving
light-cone momentum distributions of polarized nucleons in the
$^3$He nucleus.  We consider the full results at finite $Q^2$,
along with various approximations which arise in specific limits,
as well as corrections to the convolution approximation from
nucleon off-shell and non-nucleonic degrees of freedom.
Coherent effects associated with multiple scattering from two
or more nucleons in the nucleus give rise to corrections at
small values of $x$ \cite{PW99}; in this analysis we restrict
ourselves to the intermediate- and large-$x$ region, $x \gg 0$,
in which the incoherent scattering from a single nucleon is
expected to dominate.

% .......................................................................
\subsection{Weak binding approximation}
\label{ssec:wba}

A systematic framework that has been used to successfully compute
nuclear structure functions in terms of nucleon degrees of freedom
is the weak binding approximation, in which the nucleus is treated as
a nonrelativistic system of weakly bound nucleons with four-momentum
$p^\mu \equiv (M+\eps, \bm{p})$, with $|\bm{p}|, |\eps| \ll M$.
In this approach the spin-averaged $F_{1,2}$ and spin-dependent
$g_{1,2}$ structure functions of nuclei have been derived by
Kulagin {\it et al.} in Refs.~\cite{KPW, KMPW, KMd, KMhe, KP06}.
Neglecting possible nucleon structure modifications off the mass
shell (see Sec.~\ref{ssec:off} below), the spin-dependent structure
functions of $^3$He can be written, to order $\bm{p}^2/M^2$,
as \cite{KMPW, KMd, KMhe}
\begin{eqnarray}
\label{eq:convLC}
g_i^{^3{\rm He}}(x,Q^2)
&=& \int {dy \over y}
    \left[ 2 f_{ij}^p(y,\gamma)\, g_j^p\left({x\over y},Q^2\right)
           + f_{ij}^n(y,\gamma)\, g_j^n\left({x\over y},Q^2\right)
    \right],\ \ \
i,j=1,2,
\end{eqnarray}
where $y = p \cdot q / M\nu = (M + \eps + \gamma p_z)/M$ is the nuclear
light-cone momentum fraction carried by the interacting nucleon, and
a sum over indices $j$ is implied.
The functions $f_{ij}^N(y,\gamma)$ are nucleon light-cone momentum
distributions (or ``smearing functions'') in the $^3$He nucleus
computed in terms of the nuclear spectral function,
\begin{eqnarray}
\label{eq:Dy}
f_{ij}^N(y,\gamma)
&=& \int\!\!{d^4p \over (2\pi)^4}\, D_{ij}^N(\eps,\bm{p},\gamma)\
    \delta\left( y-1-\frac{\eps+\gamma p_z}{M} \right),
\end{eqnarray}
with $N = p$ or $n$.
In the Bjorken limit ($\gamma \to 1$), the smearing functions depend
only on the light-cone momentum fraction $y$, which spans the range
between $x$ and $M_{\rm He}/M \approx 3$.
At finite $Q^2$, however, they depend in addition on the variable 
$\gamma$, making them process-dependent at finite kinematics.

The energy-momentum distribution functions $D_{ij}^N$ can be
conveniently expressed in terms of coefficients of the spectral
function ${\cal P}^N$ \cite{SS},
\begin{equation}
\label{eq:spfn:2}
{\cal P}^N(\eps,\bm{p},S) = \frac12
\Big[
   {\cal F}_0^N 
 + {\cal F}_\sigma^N\, \bm{\sigma}\cdot\bm{S}
% + {\cal F}_t^N\, T_{kl}\, S_k\, \sigma_l
 + {\cal F}_t^N
   \left( \widehat{\bm{p}}\cdot\bm{S}\, \widehat{\bm{p}}\cdot\bm{\sigma}
	- \tfrac13 \bm{S}\cdot\bm{\sigma}
   \right)
% T_{kl}\, S_k\, \sigma_l
\Big],
\end{equation}
where $\widehat{\bm{p}}$ is a unit vector in the direction of $\bm{p}$,
%
%  $T_{kl} = \widehat p_k\ \widehat p_l - \tfrac13\, \delta_{kl}$
% is a traceless symmetric tensor, with
%  $\widehat p_k \equiv p_k/|\bm{p}|$,
%
and the nuclear spin vector $\bm{S}$ is defined to lie
along the $z$-axis.
The spectral coefficient ${\cal F}_0^N$ represents the spin-averaged
distribution of nucleons in the nucleus, while the spin-dependent
distributions are parametrized in terms of the longitudinal
${\cal F}_\sigma^N$ and tensor ${\cal F}_t^N$ spectral coefficients.
In general, the spectral coefficients are functions of the separation
energy $\eps$ and the magnitude $|\bm{p}|$ of the nucleon momentum,
${\cal F}_{0,\sigma,t}^N \equiv {\cal F}_{0,\sigma,t}^N(\eps,|\bm{p}|)$.
For the $g_1$ structure function the nucleon energy-momentum
distributions are given by \cite{KMPW, KMd, KMhe}
\begin{subequations}
\label{eq:Dij}
\begin{align}
D_{11}^N
&= {\cal F}_\sigma^N
 + \frac{3-\gamma^2}{6\gamma^2} \left( 3\pzhatsq - 1 \right) {\cal F}_t^N
 + \frac{p_z}{\gamma M}
   \left( {\cal F}_\sigma^N + \tfrac23 {\cal F}_t^N \right)
							\notag\\
&\ \ \ + \frac{\bm{p}^2}{M^2}
   \frac{(3-\gamma^2)\pzhatsq - 1-\gamma^2}{12\gamma^2}
   \left( 3 {\cal F}_\sigma^N - {\cal F}_t^N \right),	\label{eq:D11}\\
D_{12}^N 
&= (\gamma^2-1)
\left[
 - \frac{3\pzhatsq - 1}{2\gamma^2}\, {\cal F}_t^N
 + \frac{p_z}{\gamma M}
   \left( {\cal F_\sigma}^N
        + \left( \tfrac32\pzhatsq - \tfrac56 \right)\, {\cal F}_t^N
   \right)
\right.                                                 \notag\\
&
\left. \ \ \
    {}   
 - \frac{\bm{p}^2}{M^2}
   \left( \frac{1 + \pzhatsq (4\gamma^2-3)}{4\gamma^2}\, {\cal F}_\sigma^N
        + \frac{5 + 18\widehat{p}_z^4\gamma^2
                  - 5\pzhatsq\, (3+2\gamma^2)}{12\gamma^2}\, {\cal F}_t^N
   \right)
\right],                                                \label{eq:D12}
\end{align}
while for the $g_2$ structure function the corresponding distributions are
\begin{align}
D_{21}^N
&=
 - \frac{3\pzhatsq - 1}{2\gamma^2} {\cal F}_t^N
 - \frac{p_z}{\gamma M}
   \left( {\cal F}_\sigma^N + \tfrac23 {\cal F}_t^N \right)
 - \frac{\bm{p}^2}{M^2} \frac{3\pzhatsq - 1}{12\gamma^2}
   \left( 3 {\cal F}_\sigma^N - {\cal F}_t^N \right),	\label{eq:D21}\\
D_{22}^N
&= {\cal F}_\sigma^N
 + \frac{2\gamma^2-3}{6\gamma^2} \left( 3\pzhatsq - 1 \right) {\cal F}_t^N
 + \frac{p_z}{\gamma M}
   \left[
     (1-\gamma^2) {\cal F}_\sigma^N
   + \left( -\tfrac56 + \tfrac13 \gamma^2 + \pzhatsq (\tfrac32 - \gamma^2)
     \right) {\cal F}_t^N
     \right]                                            \notag \\
 & {} \ \ \
+ \frac{\bm{p}^2}{M^2} 
  \left[
    \frac{\pzhatsq (3 - 6\gamma^2 + 4\gamma^4)
         - 1 - 2\gamma^2}{4\gamma^2}\, {\cal F}_\sigma^N
  + \frac{5 - 2\gamma^2 (1 + 3\pzhatsq) + 4\pzhatsq\gamma^4}{12\gamma^2}
    (3\pzhatsq - 1) {\cal F}_t^N
\right].
\label{eq:D22}
\end{align}
\end{subequations}  
Note that in the $\gamma \to 1$ limit, the $D_{12}^N$ function
vanishes, in which case the nuclear $g_1$ structure function
receives contributions only from $g_1^N$.  On the other hand,
both $g_1^N$ and $g_2^N$ contribute to the nuclear $g_2$
structure function at all $Q^2$ values.
For $\gamma=1$, the diagonal functions $f_{11}^N$ and
$f_{22}^N$ integrate to the effective nucleon polarizations
(see Sec.~\ref{ssec:effpol} below), while the integral over
the off-diagonal $f_{21}^N$ smearing function vanishes.
The dependence of the smearing functions $f_{ij}^N$ on $y$ and
$\gamma$ is illustrated in Ref.~\cite{KMhe} for realistic models
of the $^3$He spectral function.

The integrated spectral function coefficient ${\cal F}_\sigma^N$
determines the average nucleon polarization in the nucleus,
\begin{equation}
\label{eq:f1norm}
\left\langle \sigma_z \right\rangle^N
= \int\!\!{d^4p \over (2\pi)^4}\, {\cal F}_\sigma^N,
\end{equation}
while ${\cal F}_t^N$ is related to the tensor polarization
\cite{SS, KMhe}.
%
% \begin{equation}
% \label{eq:f2norm}
% \left\langle T_{zk}\ \sigma_k \right\rangle^N
% = \frac29 \int\!\!{d^4p \over (2\pi)^4}\, {\cal F}_t^N.
% \end{equation}
%
For $^3$He, the integral of the function ${\cal F}_0$ gives
the number of protons (2) or neutrons (1) in the nucleus.
The average nucleon polarization can also be written in more familiar
notation in terms of the probabilities of the nucleons in the $^3$He
nucleus to be in relative $S$, $S'$ or $D$ states \cite{Friar},
\begin{subequations}
\label{eq:np-pol}
\begin{eqnarray}
\langle \sigma_z \rangle^p
&=& -\frac23\left(P_D-P_{S'}\right), \\
\langle \sigma_z \rangle^n
&=& P_S-\frac13\left(P_D - P_{S'}\right).
\end{eqnarray}
\end{subequations}
Typically, the space-symmetric $S$-state is the dominant contribution,
with the $L=0$ mixed-symmetric $S'$ state and $L=2$ tensor $D$-state
giving small corrections \cite{Friar}.

% .......................................................................
\subsection{Effective polarizations}
\label{ssec:effpol}

In the limit of zero nuclear binding and $\gamma \to 1$, the smearing
functions become infinitesimally narrow ($f_{ii}^N \sim \delta(1-y)$,
with $f_{i\not= j}^N = 0$), resulting in nuclear corrections that
are independent of $x$.
In this approximation one can express the nuclear structure functions
as linear combinations of the proton and neutron structure functions
weighted by effective polarizations $P_i^N$,
\begin{eqnarray}
\label{eq:effpol}
g_i^{^3{\rm He}}(x,Q^2)
&=& 2 P_i^p\ g_i^p(x,Q^2)\ +\ P_i^n\ g_i^n(x,Q^2),\hspace*{2cm}
i=1,2.
\end{eqnarray}
The proton effective polarizations $P_i^p$ are defined to be the
{\it average} polarizations of the two protons in the $^3$He nucleus,
rather than the total proton polarization.
The effective polarizations are defined in terms of integrals of the
diagonal smearing functions $f_{11}^N$ and $f_{22}^N$ at $\gamma=1$,
\begin{eqnarray}
P^N_i &=& \int dy\, f_{ii}^N(y,\gamma=1),
\end{eqnarray}
which can be expressed through the momentum-weighted moments
${\cal F}_m^{N (n)}$ of the spectral coefficients,
\begin{subequations}
\label{eq:effpolGEN}
\begin{eqnarray}
P^N_1
&=& {\cal F}_\sigma^{N (0)}
 - \frac13\
    \left( {\cal F}_\sigma^{N (2)} - \frac13\ {\cal F}_t^{N (2)}
    \right),						\\
P^N_2
&=& {\cal F}_\sigma^{N (0)}
 - \frac23\
   \left( {\cal F}_\sigma^{N (2)} - \frac{1}{15}\ {\cal F}_t^{N (2)}
   \right),
\end{eqnarray}
\end{subequations}
with
\begin{eqnarray}
{\cal F}_m^{N (n)}
&\equiv& \int\!\!{d^4p \over (2\pi)^4}\,
	 \left( {\bm{p} \over M} \right)^n {\cal F}_m^N(\eps,\bm{p}),
\hspace*{2cm} m = 0, \sigma, t.
\end{eqnarray}
In this notation the average nucleon polarization in
Eq.~(\ref{eq:f1norm}) can also be written as
\mbox{$\langle \sigma_z \rangle^N \equiv {\cal F}_\sigma^{N (0)}$}.
%
% $\langle T_{zk}\ \sigma_k \rangle^N \equiv (2/9) {\cal F}_t^{N (0)}$,

The effective polarizations can be computed numerically from models
of the $^3$He wave function.  Table~\ref{tab:effpol} lists values
of the coefficients in Eqs.~(\ref{eq:effpolGEN}) for the proton and
neutron obtained from the spectral function of Kievsky {\it et al.}
(KPSV) \cite{KPSV}, which is calculated using a variational approach
with a pair-correlated hyperspherical-harmonic basis.
For comparison, we also list values (in parentheses) obtained from
the spectral function of Schulze and Sauer (SS) \cite{SS}, which uses
the trinucleon bound-state wave function from Ref.~\cite{Stadler91}
computed by solving the Faddeev equations for 18 channels.

\begin{table}[tb]
\caption{Effective polarization parameters ${\cal F}_\sigma^{N(0)}$,
	${\cal F}_\sigma^{N(2)}$, ${\cal F}_t^{N(2)}$ and the average
	polarizations $P_1^N$ and $P_2^N$ for the neutron and proton,
	from the KPSV \cite{KPSV} and SS \cite{SS} (in parentheses)
	spectral functions.}
\centering
\begin{tabular}[c]{l c c c c c}  \hline
\hspace{0.5cm}
  &\ \ ${\cal F}_\sigma^{N(0)}$\ \
  &\ \ ${\cal F}_\sigma^{N(2)}$\ \
  &\ \ ${\cal F}_t^{N(2)}$\ \ 
  &\ \ \ \ $P_1^N$\ \ &\ \ $P_2^N$\ \			\\ \hline
neutron \hspace{0.5cm}
  & 0.856         & 0.018      & 0.013\ \   
  &\ \ 0.851      & 0.844				\\
  & (0.888)       & (0.016)    & (0.010)\ \
  &\ \ (0.884)    & (0.878)				\\ \hline
proton  \hspace{0.5cm}
  & $-0.029$      & $-0.002$   & 0.009\ \
  &\ \ $-0.028$   & $-0.028$				\\
  & $(-0.022)$    & $(-0.001)$ & (0.004)\ \
  &\ \ $(-0.021)$ & $(-0.021)$				\\ \hline
\end{tabular}
\label{tab:effpol}
\end{table}

The lowest order neutron coefficient $\langle \sigma_z \rangle^n$
dominates all other contributions, giving an average neutron
polarization of $\approx 86\%$ (89\%) for the KPSV (SS) spectral
function.
The small negative value of the average proton polarization
reflects the preferential antialignment of the spins of the
proton pair, with $\langle \sigma_z \rangle^p \approx -3\% (-2\%)$
for the two models.
Other models, such as the PEST three-body wave function
\cite{Bissey01}, give similar values,
$\langle \sigma_z \rangle^n = 88\%$ and
$\langle \sigma_z \rangle^p = -2\%$, as does an earlier world average
of three-nucleon models,
$\langle \sigma_z \rangle^n = 86 \pm 2\%$ and
$\langle \sigma_z \rangle^p = -2.8 \pm 0.4\%$ \cite{Friar}.

The $\bm{p}^2$-weighted moment ${\cal F}_\sigma^{N(2)}$ is
$\approx 2\%$ of the average polarization for the neutron and
$\approx 3-4\%$ for the proton.  The $\bm{p}^2$-weighted tensor
moment ${\cal F}_t^{N(2)}$ is $\approx 1\%-1.5\%$ of the leading
${\cal F}_\sigma^{N(0)}$ term for the neutron, but a somewhat
larger fraction, $\approx 10\%-15\%$, for the proton, and with
opposite sign (although, as noted, the proton average polarization
itself is very small).
In practice, the additional suppression factors of $\sim 10$ and
$\sim 20$ for the $P_1^N$ and $P_2^N$ effective polarization in
Eqs.~(\ref{eq:effpolGEN}), respectively, render the tensor
contributions negligible.  Overall, the higher order coefficients
reduce the magnitude of the neutron polarization by
$\approx 1\%-1.5\%$, and the proton polarization by
$\approx 2\%-3\%$.
On the scale of the nuclear wave function model dependence
\cite{KPSV, SS} of the effective polarizations, which amounts
to $\sim 4\%$ for the neutron and $\sim 15\%$ for the proton,
the higher order corrections are not significant.

% .......................................................................
\subsection{Non-nucleonic contributions}
\label{ssec:Delta} 

While scattering from nucleons in the nucleus gives the dominant
contribution to nuclear DIS, there are indications that a description
of nuclear properties in terms of nucleon degrees of freedom alone
may not be complete.  Pions and vector mesons have long been recognized
as playing an important role in the structure and interactions of
nucleons at low energies, and their effects may also be relevant
in high energy reactions such as DIS.  A notable example is the
nuclear EMC effect, or the ratio of nuclear to deuteron structure
functions, which deviates from unity due to the redistribution of
momentum between nucleons and pions in the nucleus \cite{Ericson83}.
DIS from pions and other mesons exchanged between different nucleons
in the nucleus can also lead to antishadowing effects at $x \sim 0.1$
in unpolarized structure functions \cite{K_MEC, KU_MEC, MT_MEC}.
More recently, it was observed \cite{Cloet09} that the presence of an
isovector-vector $\rho^0$ mean field in asymmetric nuclei can induce
a shift in the $u$ and $d$ quark distributions which has important
consequences for the NuTeV anomaly.

For spin-dependent observables, a small admixture of the $\Delta(1232)$
isobar in the three-body wave function \cite{Saito90} was found to be
necessary to understand the experimental value of the axial vector
charge measured in $^3$H $\beta$ decay \cite{Budick91}.  The same
mechanism was argued \cite{Frank96, Boros01} to contribute also to
the isovector $g_1$ structure function for the $^3$He--$^3$H system,
whose lowest moment is given by the Bjorken sum rule \cite{Bj66}
\begin{eqnarray}
\label{eq:GamRatio}
{\Gamma_1^{^3{\rm H}(1)} - \Gamma_1^{^3{\rm He}(1)}\
 \over \Gamma_1^{p(1)} - \Gamma_1^{n(1)}}\
 =\ {g_A^{^3{\rm H}} \over g_A}.
\end{eqnarray}
Experimentally, one finds an $\approx 4\%$ suppression of the
axial vector charge for $^3$H compared with the nucleon,
$g_A^{^3{\rm H}} / g_A = 0.956 \pm 0.004$ \cite{Budick91}.
Neglecting the Fermi motion of the $\Delta$ baryon in the nucleus,
the $\Delta$ contribution to the nuclear $g_1$ structure function
was incorporated by Bissey {\it et al.} \cite{Bissey01} in terms
of off-diagonal $N \to \Delta$ transition structure functions and
corresponding effective polarizations $P_1^{N\Delta}$.
For a $^3$He target, the total $g_1$ structure function is then
given by
\begin{eqnarray}
\label{eq:g1withDelta}
g_1^{^3{\rm He}}(x,Q^2)
&=& g_1^{^3{\rm He}}(x,Q^2)\Big|_N\
 +\ g_1^{^3{\rm He}}(x,Q^2)\Big|_\Delta,
\end{eqnarray}
where the nucleonic contribution $g_1^{^3{\rm He}}(x,Q^2)\big|_N$
is given by Eq.~(\ref{eq:convLC}) (or the effective polarization
approximation, Eq.~(\ref{eq:effpol})), and
\begin{eqnarray}
\label{eq:effpolwithD}
g_1^{^3{\rm He}}(x,Q^2)\Big|_\Delta
&=& 2 \left[ P_1^{\, n\, \Delta^0}\, g_1^{\, n\, \Delta^0}(x,Q^2)\
	  +\ P_1^{\, p\, \Delta^+}\, g_1^{\, p\, \Delta^+}(x,Q^2)
      \right].
\end{eqnarray}
In valence quark models the nucleon $g_1$ structure function can be
decomposed into contributions involving scalar and axial vector
spectator diquarks \cite{CT88, Boros99}, which allows the transition
structure functions to be related by \cite{Bissey01}
\begin{eqnarray}
\label{eq:newSF}
g_1^{\, n\, \Delta^0}(x,Q^2)\ =\ g_1^{\, p\, \Delta^+}(x,Q^2)
&=& {2 \sqrt{2} \over 5}
    \left[ g_1^p(x,Q^2) - 4 g_1^n(x,Q^2) \right].
\end{eqnarray}
The effective transition polarizations can then be determined
from Eqs.~(\ref{eq:GamRatio})--(\ref{eq:newSF}) in terms of the
diagonal polarizations $P_1^N$ and the moments $\Gamma_1^{N(1)}$
of the nucleon $g_1$ structure functions,
\begin{eqnarray}
\label{eq:effNDelPol}
P_1^{\, n\, \Delta^0} + P_1^{\, p\, \Delta^+}
&=& {5 \over 4 \sqrt{2}}
    { (P_1^n - P_1^p - g_A^{^3{\rm H}}/g_A)
      (\Gamma_1^{p(1)} - \Gamma_1^{n(1)})
     \over \Gamma_1^{p(1)} - 4\Gamma_1^{n(1)} }.
\end{eqnarray}
Using the most recent de~Florian {\it et al.} (DSSV) \cite{DSSV09}
parametrization of the spin-dependent parton distribution functions
(PDFs) at $Q^2 = 5$~GeV$^2$ and the KPSV values for the effective
polarizations \cite{KPSV}, we find
$P_1^{\, n\, \Delta^0} + P_1^{\, p\, \Delta^+} = -0.0125$.
This can be compared with the value $-0.012$ obtained by Bissey
{\it et al.} \cite{Bissey01} using the earlier GRSV spin-dependent
PDFs \cite{GRSV01}, and including corrections from nucleon off-shell
effects \cite{Steffens99} (see Sec.~\ref{ssec:off} below).

For the $g_2$ structure function, there is no corresponding isovector
sum rule analogous to the Bjorken sum rule.  However, in the WW
(leading twist) approximation, and in the absence of nuclear Fermi
motion, the $\Delta$ contributions to $g_2^{^3{\rm He}}$ can be
expressed in analogy with those for $g_1^{^3{\rm He}}$ in
Eq.~(\ref{eq:effpolwithD}), which we shall use in our numerical
estimates in Sec.~\ref{sec:results}.

% .......................................................................
\subsection{Nucleon off-shell corrections}
\label{ssec:off} 

In the derivation of the one-dimensional convolution representation
of the nuclear structure function in Eq.~(\ref{eq:convLC}), the
partonic structure of the free nucleon was assumed to be unaltered
when the nucleon is placed inside the nucleus.
Off-shell dependence of the nucleon structure functions would
require a generalization of (\ref{eq:convLC}) to take into account
the additional dependence of $g_{1,2}^N$ on $p^2 \not= M^2$.
In the WBA, to order $\bm{p}^2/M^2$ in the nucleon momentum, one can
write a generalized, two-dimensional convolution for the nuclear
structure function in terms of a $y$- and $p^2$-dependent smearing
function and a ($p^2$-dependent) off-shell nucleon structure function
\cite{KMPW}
\begin{eqnarray}
\label{eq:2Dconv}
g_i^{^3{\rm He}}(x,Q^2)
&=& \int {dy \over y} \int dp^2
    \left[
	2 \widetilde{f}_{ij}^p(y,\gamma,p^2)\,
	  g_j^p\left({x\over y},Q^2,p^2\right)
	+ \widetilde{f}_{ij}^n(y,\gamma,p^2)\,
	  g_j^n\left({x\over y},Q^2,p^2\right)
    \right],		\nonumber\\
& &
\end{eqnarray}
where $i,j=1,2$ and the off-shell dependent smearing function is
defined in analogy with that in Eq.~(\ref{eq:Dy}) by
\begin{eqnarray}
\label{eq:Dyp2}
\widetilde{f}_{ij}^N(y,\gamma,p^2)
&=& \int\!\!{d^4p \over (2\pi)^4}\, D_{ij}^N(\eps,\bm{p},\gamma)\
    \delta\left( y-1-\frac{\eps+\gamma p_z}{M} \right)
    \delta\left( p^2 - (M+\eps)^2 + \bm{p}^2 \right).
%    \delta\left( p^2 - M^2 - 2M (\eps - \bm{p}^2/2M) \right).
\end{eqnarray}
The presence of the $p^2$ dependence in the nucleon structure functions
in Eq.~(\ref{eq:2Dconv}) does not allow the $y$ and $p^2$ integrations
to decouple, as in the on-shell convolution expression (\ref{eq:convLC}).
Note that from Eqs.~(\ref{eq:Dy}) and (\ref{eq:Dyp2}), the smearing
function $f_{ij}^N(y,\gamma)$ is obtained simply by integrating the
off-shell function $\widetilde{f}_{ij}^N(y,p^2,\gamma)$ over $p^2$,
\begin{eqnarray}
\label{eq:fyp2} 
f_{ij}^N(y,\gamma)
&=& \int dp^2\, \widetilde{f}_{ij}^N(y,\gamma,p^2).
\end{eqnarray}

The dependence of the bound nucleon structure function on the
off-shell mass $p^2$ is generally difficult to determine.
In fact, the concept of nucleon off-shell effects is inherently a
theoretical construct which is strictly defined only within a
specific definition of the nucleon fields; field redefinitions can
in principle be made to move strength between wave function
and off-shell contributions, with only the total structure function
being physical.  For the case of the deuteron structure function,
this was demonstrated for both unpolarized and polarized scattering
in Refs.~\cite{MST94, MPT, PMT} in a simple spectator quark model.
A number of other models have also been considered in attempts to
quantify the possible modification of the nucleon substructure in the
nuclear medium \cite{KP06, Steffens99, GL92, MSS97, Cloet05, CJ11,
CJ12}.

In the WBA, the average nuclear binding and kinetic energies of
the in-medium nucleons are small compared with the nucleon mass,
so that the typical nucleon virtuality is
\mbox{$|p^2 - M^2|/M^2 \ll 1$}.
In this case, the bound nucleon structure function can be expanded
in a Taylor series about the on-shell limit \cite{KMPW},
\begin{eqnarray}
\label{eq:g1off}
g_i^N(x,Q^2,p^2)
&=& g_i^N(x,Q^2)
    + (p^2-M^2) \left. {\partial g_i^N \over \partial p^2}
		\right|_{p^2=M^2}.
\end{eqnarray}
To determine the $p^2$ derivative of the off-shell structure functions,
we take the leading twist approximation for $g_1^N$ and $g_2^N$, and
assume that the spin-dependent quark distribution at a scale
$Q^2 = Q_0^2$ can be written in the form of a spectral representation,
\begin{eqnarray}
\label{eq:Delq_off}
\Delta q(x,p^2)
&=& \int ds \int_{-\infty}^{k^2_{\rm max}(x,p^2)} dk^2\,
    \rho(s,x,p^2,k^2),
\end{eqnarray}
where $k$ is the four-momentum of the interacting quark,
with maximum virtuality $k^2_{\rm max} = x(p^2 - s/(1-x))$,
and $s = (p-k)^2$ is the invariant mass squared of the
spectator quark system.
Since our focus is mainly on the nuclear effects in the large-$x$
region, we consider the application of the model specifically to
the spin-dependent valence quark PDFs.  In this case the quark
spectral function may be approximation by a single pole at mass
$s=s_0$ \cite{KP06},
\begin{eqnarray}
\label{eq:rho_q}
\rho &\to& \delta(s-s_0)\, \Phi(k^2, p^2).
\end{eqnarray}
Fits to the free nucleon structure functions in the valence quark
region suggest values of the invariant spectator masses squared
$\sim 2$~GeV$^2$ \cite{KP06}, and in practice we use
$s_0 = 2.1$~GeV$^2$.
Following Refs.~\cite{KP06, CJ11}, the off-shell dependence of
the quark spectral function can be parametrized through the $p^2$
dependence of the ultraviolet cutoff parameter $\Lambda_N(p^2)$
used to regulate the $k^2$ integration in Eq.~(\ref{eq:Delq_off}).
The cutoff $\Lambda_N$ can be related to the radius of confinement
of the nucleon $R_N$, $\Lambda_N \sim 1/R_N$, with the variation with
$p^2$ reflecting the amount of nucleon swelling in the nuclear medium.
From the analysis of the nuclear EMC effect in the $Q^2$ rescaling
model \cite{Close85}, typical values for nucleon swelling in the
$^3$He nucleus were found to be $\delta R_N/R_N \approx 4.0\%-4.7\%$.

Within this framework, the $p^2$ derivative of the spin-dependent
structure function $g_i^N$ becomes
\begin{eqnarray}
\label{eq:dqdp2}
M^2 \left. {\partial g_i^N \over \partial p^2}
    \right|_{p^2=M^2}
&=& c_N\, g_i^N\ +\ h_N(x)\, {\partial g_i^N \over \partial x},
\end{eqnarray}
where
\begin{eqnarray}
\label{eq:h(x)}
h_N(x) &=& x(1-x) \frac{ (1-\lambda_N)(1-x) M^2 + \lambda_N s_0 }
                     { (1-x)^2 M^2 - s_0 }.
\end{eqnarray}
The scale parameter $\lambda_N$ is the $p^2$ derivative of the cutoff
$\Lambda_N$, and can be expressed in terms of the change in confinement
scale $\delta R_N/R_N$ and the average virtuality of the nucleon
$\langle \delta p^2 \rangle_N$,
\begin{eqnarray}
\lambda_N
&\equiv& \left. {\partial \ln \Lambda_N^2 \over \partial \ln p^2} 
         \right|_{p^2=M^2}
= -2\, {\delta R_N / R_N \over
        \langle \delta p^2 \rangle_N / M^2},
\label{eq:lamndaN}
\end{eqnarray}
where
\begin{eqnarray}
\langle \delta p^2 \rangle_N
&=& \int dy\, dp^2\, (p^2-M^2)\, \widetilde{f}_0^N(y,p^2,\gamma). 
\end{eqnarray}
Here the function $\widetilde{f}_0^N$ is the spin-averaged
analog of the off-shell nucleon smearing functions in $^3$He.
Note that because the proton and neutron momentum distributions in
$^3$He are not identical, the average value of the virtuality of
the bound proton and neutron in $^3$He will in general be different.
Using the KPSV spectral function, the average virtualities for the
proton and neutron are 
$\langle \delta p^2 \rangle_n/M^2 \approx -9.5\%$ and
$\langle \delta p^2 \rangle_p/M^2 \approx -7.2\%$, respectively.
(For comparison, the corresponding average virtuality
of nucleons in a deuterium nucleus is $\approx -(4-6)\%$
\cite{CJ11}.)

For spin-averaged PDFs, the normalization coefficient $c_N$ in
Eq.~(\ref{eq:dqdp2}) was computed in Ref.~\cite{CJ11} by requiring
that the off-shell corrections do not modify the valence quark
number, while Kulagin and Petti \cite{KP06} imposed the sum of the
off-shell and shadowing corrections to not renormalize the valence
quark number.  For the axial vector current, there is no corresponding
sum rule for a bound nucleon; however, it is reasonable to assume  
a similar invariance of the axial charge for a given flavor $q$ in
the nuclear medium.  As discussed in Sec.~\ref{ssec:Delta} above,
the axial vector charge in the $\beta$ decay of $^3$H differs by
$\approx 4\%$ from that in neutron $\beta$ decay, and some of
this difference could arise from nucleon off-shell corrections
\cite{Bissey01}.
In the framework of the present analysis, the entire difference is
attributed to wave function effects and corrections arising from
$\Delta$ components in the nuclear wave function \cite{Saito90},
so that off-shell corrections do not modify the lowest moments of the
valence quark PDFs.  This assumption is equivalent to the condition
\begin{eqnarray}
\int_0^1 dx\,
\left. {\partial g_i^N \over \partial p^2} \right|_{p^2=M^2}
&=& 0,
\label{eq:norm}
\end{eqnarray}
which leads to the constraint on the normalization constant
\begin{eqnarray}
c_N &=& - {\int_0^1 dx\, h_N(x)\, (\partial g_i^N/\partial x)
	   \over \Gamma_i^{N(1)}}.
\label{eq:c}
\end{eqnarray}
In the following we will refer to this model as the off-shell covariant
spectator (OCS) model.

In a somewhat different framework, Steffens {\it et al.}
\cite{Steffens99} used the quark-meson coupling (QMC) model to compute
the effects of mean field potentials in the nucleus using the local
density approximation.  The in-medium scalar ($\sigma$) and vector
($\rho$, $\omega$) fields modify the quark-meson couplings, inducing
changes in the nucleon's mass and energy, as well as the energy of
intermediate state.  For the quark distributions in the free nucleon
the MIT bag model was used, which restricts the validity of the
calculation to $0.2 \lesssim x \lesssim 0.7$.  The net effect is a
small, flavor-dependent correction, which was parametrized in terms
of the ratio of the PDFs in the free and bound nucleons,
\begin{eqnarray}
\label{eq:offQMC}
{ \Delta q_v(x) \over \Delta \widetilde{q}_v(x) }
&=& a_q\, x^{b_q}\ +\ c_q\, x^{d_q} (1-x)^{e_q},
\end{eqnarray}
with the parameters $a_q, \cdots, e_q$ given in Ref.~\cite{Steffens99}
for the $^3$He and $^6$Li nuclei.
In the next section we compare the effects of these models on the
$g_1^{^3{\rm He}}$ structure function and estimate the uncertainty
arising from the off-shell model dependence.

%%%%%%%%%%%%%%%%%%%%%%%%%%%%%%%%%%%%%%%%%%%%%%%%%%%%%%%%%%%%%%%%%%%%%%%%%
\section{Numerical results}
\label{sec:results}

In this section we present the numerical results for the various
spin-dependent nuclear corrections described in Sec.~\ref{sec:nuclear}.
We consider the effects on the $x$ dependence of the $g_1$ and $g_2$
structure functions of $^3$He, and several of their low moments,
as well as on the polarization asymmetries that are more directly
accessed in inclusive scattering experiments.
In particular, we study the impact of the different corrections and
their uncertainties on the extraction of the spin structure of the
neutron, and the accuracy of the various approximations used in the
literature.
We examine the corrections in both the DIS and nucleon resonance
regions, using, for illustration, the DSSV \cite{DSSV09} leading
twist parametrization of the spin-dependent PDFs in the former,
and the MAID parametrization for the low-$W$ region.
To estimate the dependence of the results on the input nucleon
parametrization, we also compare with the parametrization from
Ref.~\cite{Simula02}.
For the nucleon distributions in $^3$He, we use the spin-dependent
KPSV spectral function \cite{KPSV}, but consider also the results
with the SS \cite{SS} spectral function.

% .......................................................................
\subsection{Structure Functions}
\label{ssec:sfs} 

To begin our discussion of the nuclear effects on the $^3$He
structure functions and polarization asymmetries, we note that
since the latter are ratios of spin-averaged to spin-dependent
structure functions, they will in general also depend on the nuclear
corrections in the unpolarized $F_1$ and $F_2$ structure functions.
To determine the role played by nuclear effects in the asymmetries,
it is therefore necessary to first understand the corrections to the
$^3$He $F_1$ and $F_2$ structure functions.

\begin{figure}[bt]
\includegraphics[width=8cm]{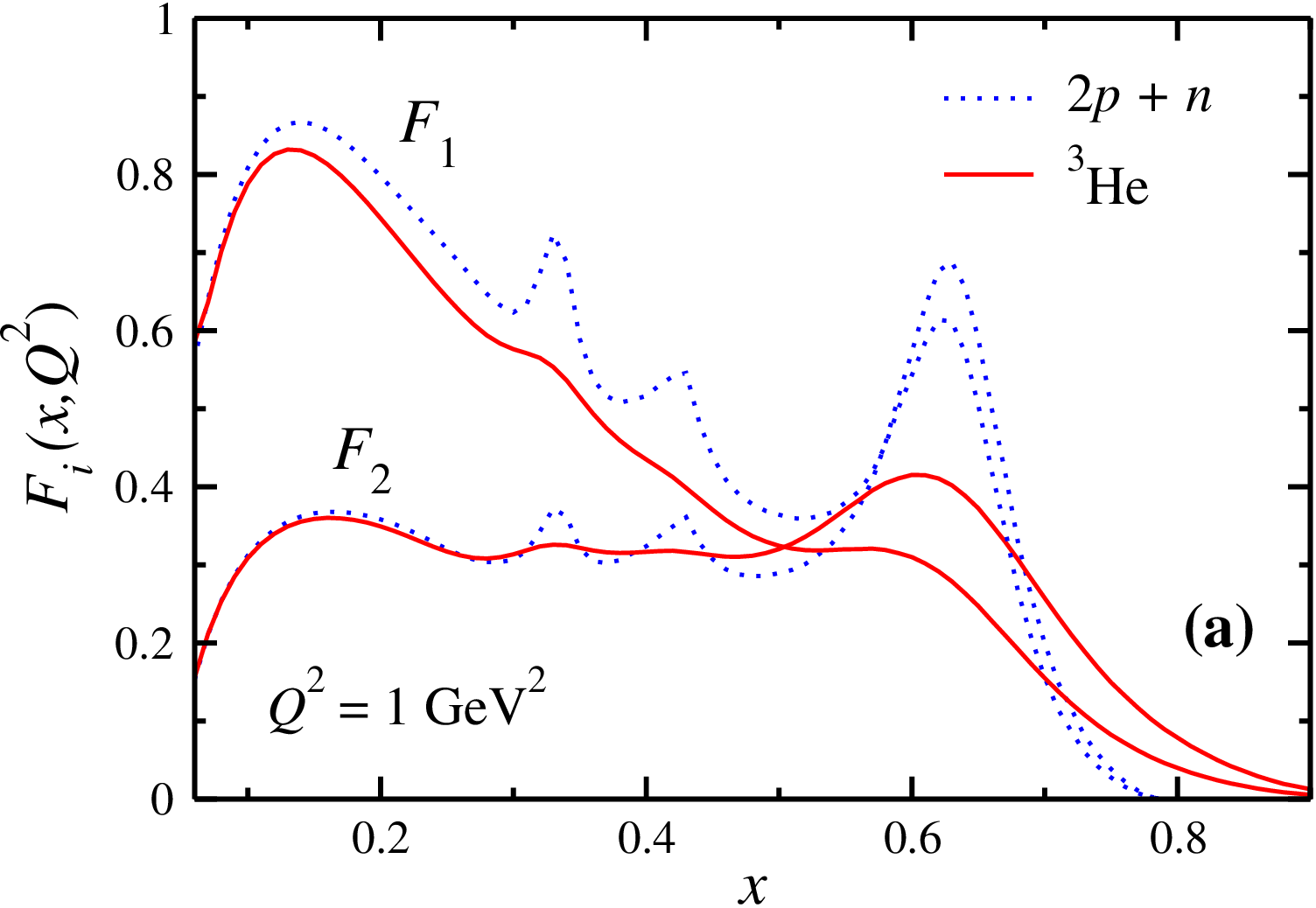}\ \ \
\includegraphics[width=8cm]{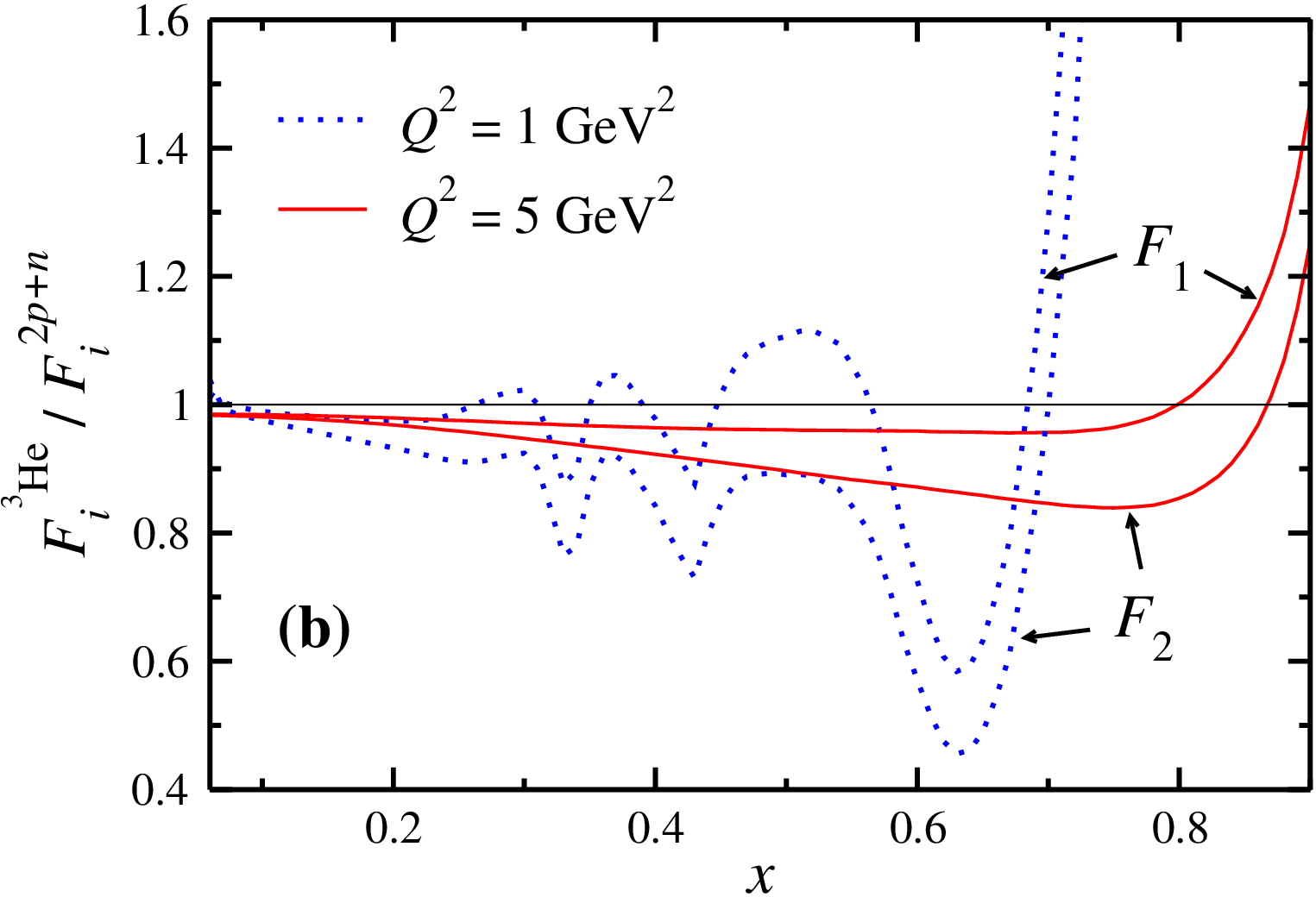}
\caption{{\bf (a)}
	Spin-averaged $F_1$ and $F_2$ structure functions
	of $^3$He (solid) and $(2p+n)$ (dotted), using the
	Bosted-Christy nucleon structure function parametrization
	\cite{BC} at $Q^2 = 1$~GeV$^2$.
	{\bf (b)}
	Ratios of $F_1$ and $F_2$ structure functions of
	$^3$He to $(2p+n)$ at $Q^2 = 1$~GeV$^2$ (dotted)
	and $Q^2 = 5$~GeV$^2$ (solid), using the BC \cite{BC} and
	NMC \cite{NMC} nucleon structure function parametrization
	for the resonance and DIS regions, respectively.
	The $^3$He structure functions in all cases are computed
	using the convolution formalism including finite $Q^2$
	corrections \cite{KP06}.}
\label{fig1}
\end{figure}

In Fig.~\ref{fig1} the spin-averaged $F_1$ and $F_2$ structure
functions for $^3$He are compared with those for the corresponding
nucleon isospin combination, $2p+n$, at $Q^2 = 1$ and 5~GeV$^2$.
Note that at finite $Q^2$ the nuclear $F_1$ structure function receives
contributions from the nucleon $F_1$ and $F_2$ structure functions,
while the nuclear $F_2$ structure functions depends only on $F_2^N$
at any $Q^2$.
At the lower $Q^2$ value, the resonance structures are prominent
at large values of $x$ for the free proton and neutron structure
functions, particularly in the region of the $\Delta(1232)$ resonance.
For the input nucleon $F_1$ and $F_2$ structure functions we use
the resonance parametrization of Bosted and Christy \cite{BC}.
After applying the nuclear smearing corrections in the WBA, using the
spin-averaged analogs \cite{KP06} of the nucleon momentum distribution
functions $f_{ij}^N$ in Sec.~\ref{ssec:wba} with the KPSV spectral
functions, the resonance peaks are significantly smeared out.
The effects is strongest in the $\Delta$ region, where the smearing
reduces the height of the peaks by a factor of $\approx 2$.

This is also evident in the ratio of the $^3$He to $2p+n$ structure
functions shown in Fig.~\ref{fig1}(b), where the resonance structures
are effectively inverted compared to those in the $F_1$ and $F_2$
functions themselves.
In contrast, the ratios of the $^3$He to nucleon structure functions
in the deep-inelastic region at $Q^2 = 5$~GeV$^2$, computed using the
NMC parametrization \cite{NMC}, show the smooth behavior characteristic
of the nuclear EMC effect, with a depletion at $x \sim 0.7$ and a
subsequent rise above unity at larger $x$ due to Fermi motion.

\begin{figure}
\includegraphics[width=8cm]{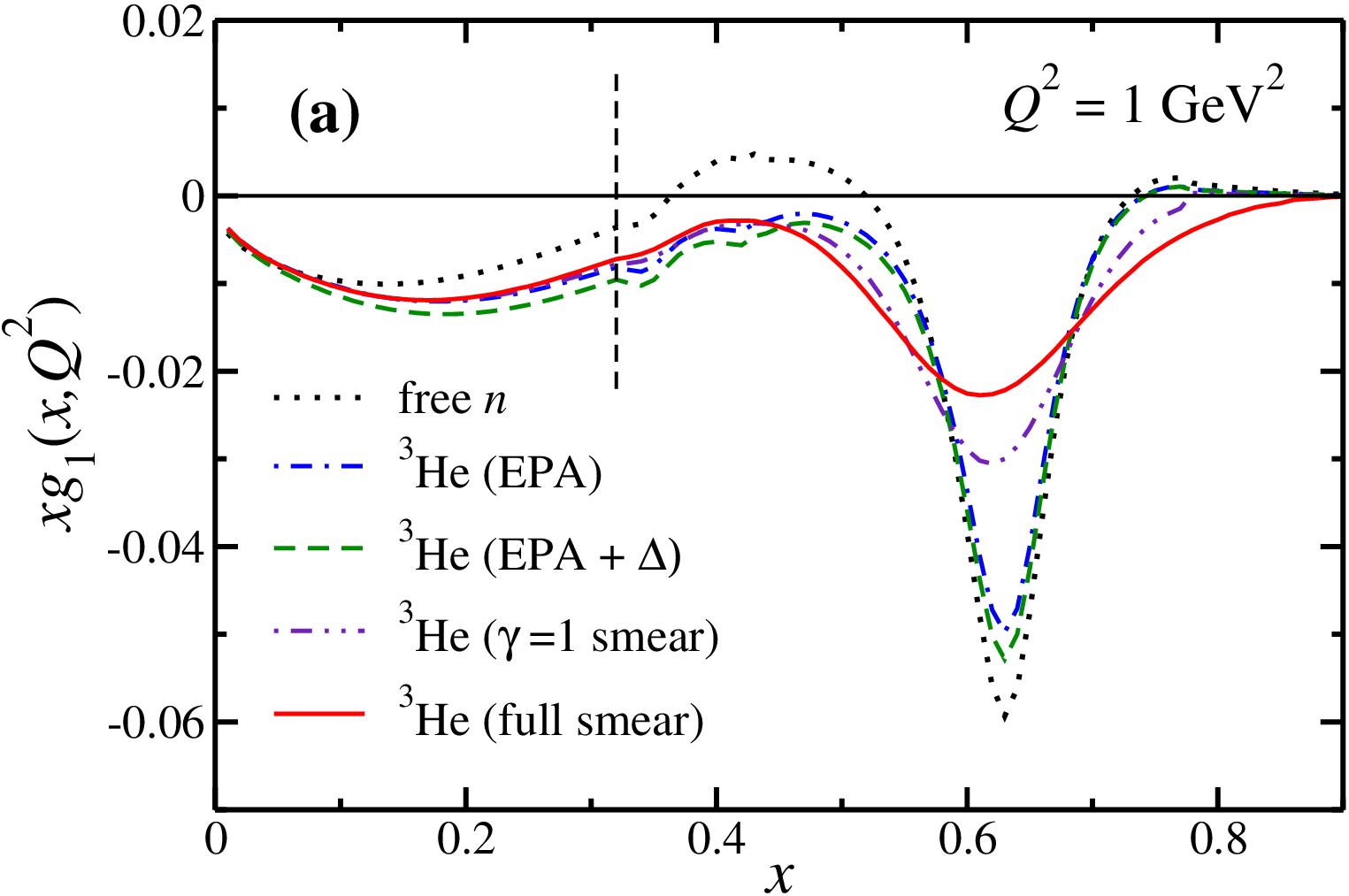}
\includegraphics[width=8cm]{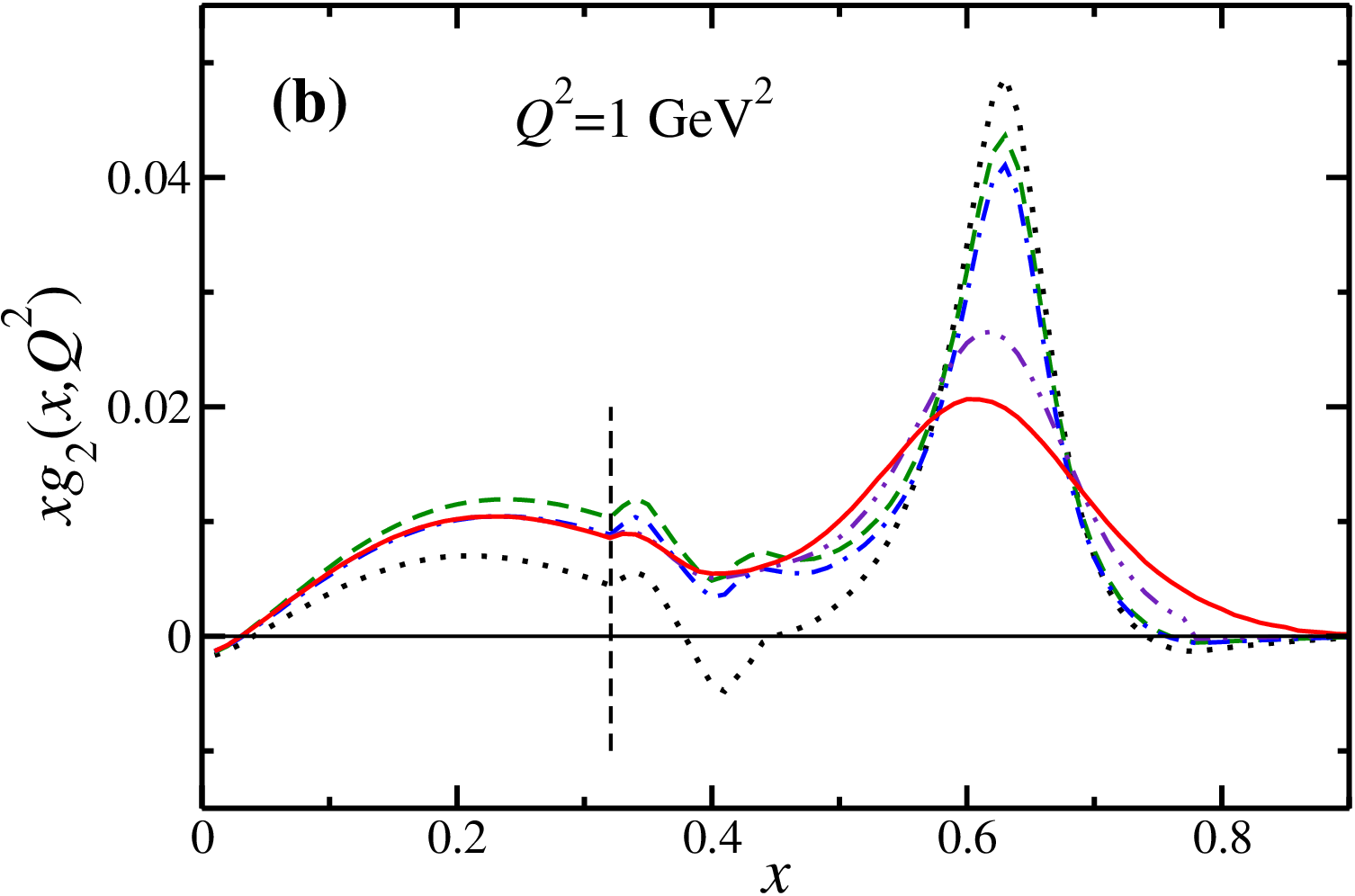}\\
\vspace*{0.9cm}
\includegraphics[width=8cm]{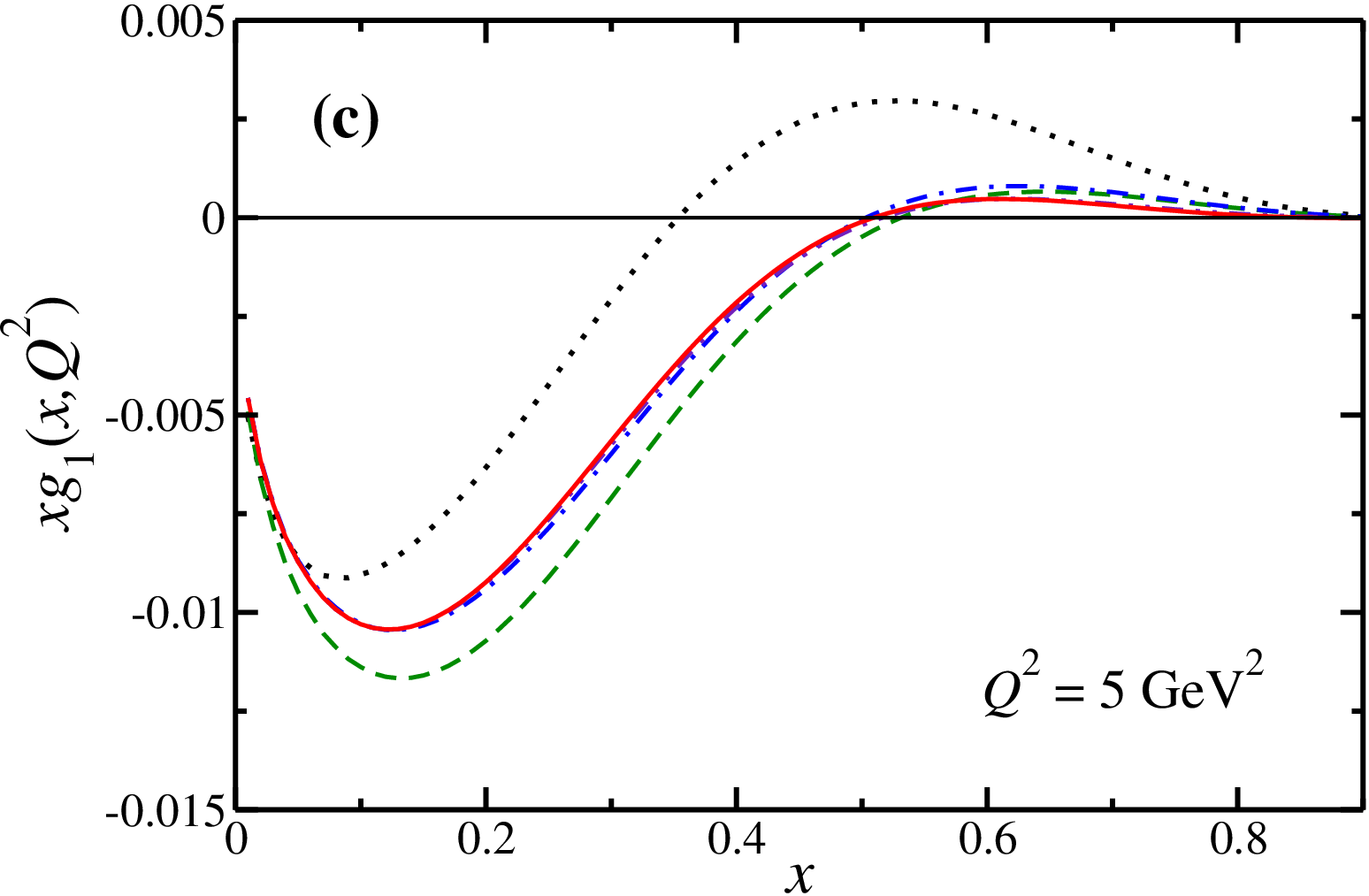}
\includegraphics[width=8cm]{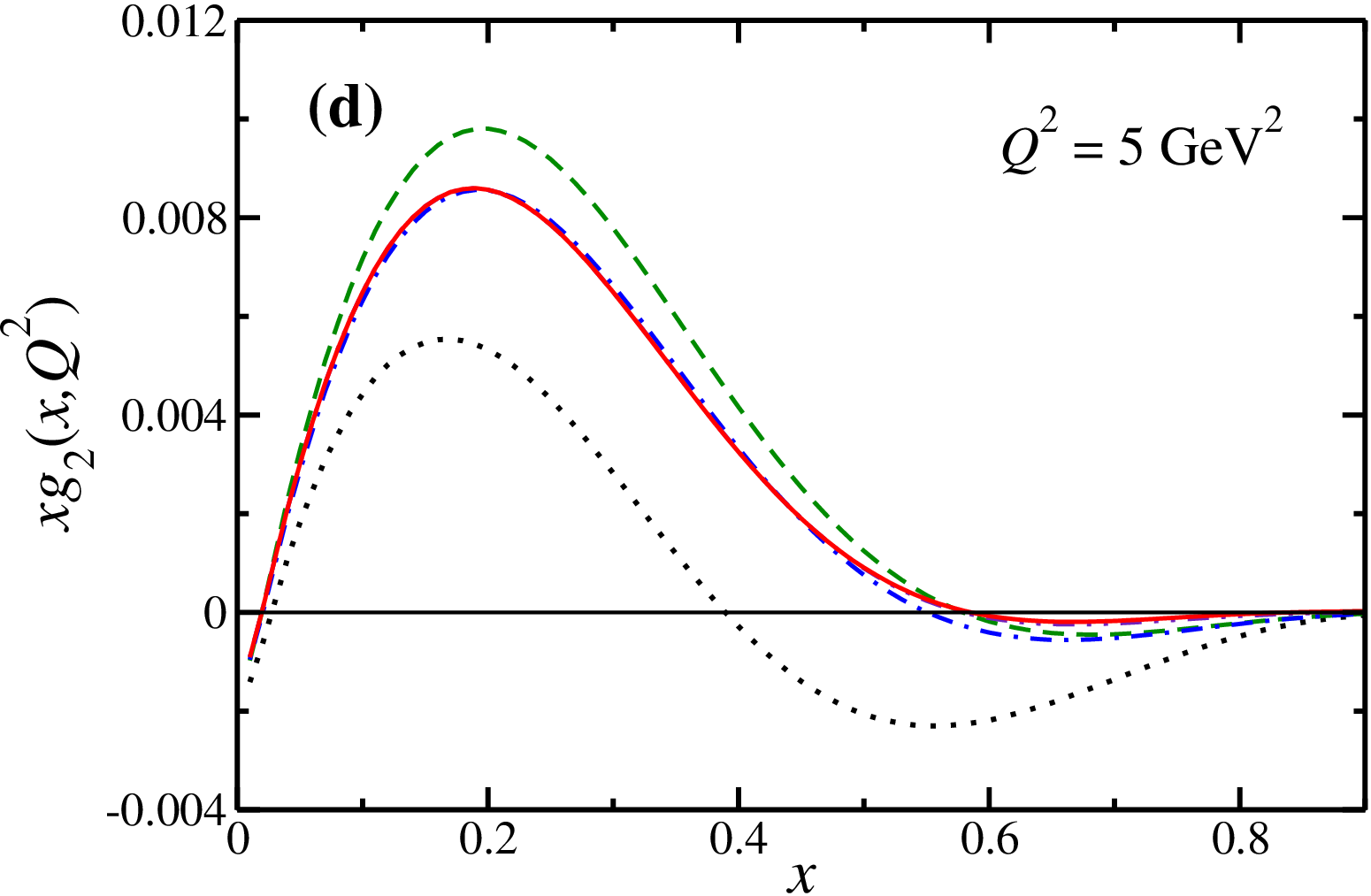}
\caption{Spin-dependent $xg_1$ and $xg_2$ structure functions of the
	neutron (black dotted) and $^3$He, computed in the EPA with
	nucleon only (blue dot-dashed) and with $\Delta$ components
	(green dashed), and with Fermi smearing in the Bjorken
	($\gamma=1$) limit (violet dot-dot-dashed) and at finite
	$Q^2$ (red solid).
	The functions at $Q^2 = 1$~GeV$^2$
		[{\bf (a)} and {\bf (b)}]
	are computed from the MAID \cite{MAID99} and DSSV \cite{DSSV09}
	parametrizations of the nucleon resonance and DIS regions,
	respectively (the dashed vertical line indicating the
	boundary between these), while those at $Q^2 = 5$~GeV$^2$
		[{\bf (c)} and {\bf (d)}]
	use the DSSV fit.}
\label{fig2}
\end{figure}

Qualitatively similar effects of smearing are observed for the
spin-dependent $^3$He structure functions $g_1$ and $g_2$ in
Fig.~\ref{fig2}.  Here the functions $xg_1$ and $xg_2$ for the
free neutron are compared with the corresponding $^3$He functions
computed using the various approximations discussed in
Sec.~\ref{sec:nuclear}, both in the resonance region at
$Q^2 = 1$~GeV$^2$ (Figs.~\ref{fig2}(a) and (b)) and in the
DIS region at $Q^2 = 5$~GeV$^2$ (Figs.~\ref{fig2}(c) and (d)).
In particular, we compute the $^3$He structure functions using the
effective polarization approximation (EPA), Eq.~(\ref{eq:effpol}),
including the effects of $\Delta$ components in the nuclear
wave function, Eq.~(\ref{eq:effpolwithD}), and accounting for
Fermi smearing effects, both in the Bjorken limit ($\gamma=1$)
and at finite $Q^2$, Eq.~(\ref{eq:2Dconv}).

For the input nucleon structure functions we use the MAID model
\cite{MAID99} for the resonance region at low $W$ values, and
the DSSV leading twist parametrization \cite{DSSV09} in the
DIS region, which is taken here to be $W^2 \geq 3$~GeV$^2$.
At $Q^2 = 1$~GeV$^2$, the boundary between these (indicated
by the dashed vertical lines in Figs.~\ref{fig2}(a) and (b)),
occurs at $x \approx 0.32$.
At this $Q^2$ the dominant feature in the structure functions
is the strong $\Delta$ resonance peak at $x \approx 0.6$.
Compared with the free neutron, the $\Delta$ peak in the $^3$He
structure functions computed from the full smearing function
in Eq.~(\ref{eq:Dy}) is reduced by more than a factor 2.
As noted in Ref.~\cite{KMhe} and illustrated in Fig.~\ref{fig2}(a)
and (b), using the smearing function computed in the Bjorken limit
underestimates the amount of smearing, with the $\Delta$ peak in
the $^3$He structure function some 20\%--30\% larger in magnitude
than for the full, $Q^2$-dependent smearing.
The difference between the smeared results and those obtained
from the EPA are even more striking, with the EPA reducing the
neutron structure functions by only a few percent.  The addition
of the $\Delta$ contribution increases the magnitude of the
functions slightly, but in either case it is clear that the
approximation of $x$-independent nuclear corrections breaks down
in the region where the structure is dominated by resonances.

The EPA approach is expected to be more reliable in the DIS region
at high $W$, where the structure functions are considerably smoother.
The corresponding $xg_1$ and $xg_2$ functions are illustrated in
Fig.~\ref{fig2}(c) and (d) at $Q^2 = 5$~GeV$^2$.  Here the resonance
structure at low $W$ is restricted to larger $x$, and for most $x$
values the functions are dominated by the nonresonant continuum,
so that it is reasonable to approximate $g_1$ and $g_2$ by the
leading twist contributions \cite{DSSV09}.
Away from the $x \sim 1$ region, where smearing effects will come
into play, one can understand the relative differences between
the neutron and $^3$He structure functions simply within the EPA.
From Eq.~(\ref{eq:effpol}) and Table~\ref{tab:effpol}, the neutron
effective polarization $P_1^n$ reduces the magnitude of the (negative)
$g_1^n$ structure function by $\approx 15\%$.  However, the proton
contribution, which is given by the product of the small (negative)
effective polarization, $2 P_1^p \approx -5\%$, and the large (positive)
$g_1^p$ structure function, shifts the overall $^3$He structure
function downward, rendering $g_1^{^3{\rm He}} < g_1^n$.
This is seen in the comparison in Fig.~\ref{fig2}(c) and in the
DIS region at small $x$ in Fig.~\ref{fig2}(a).

The effects of nuclear smearing in the DIS region, either for
$\gamma=1$ or including the finite-$Q^2$ corrections, are negligible
at $x \lesssim 0.5$ compared with the EPA with nucleons.
At larger $x$ the smearing effects are more significant, although the
magnitude of the structure functions there is considerably smaller.
(Note that because the neutron structure function changes sign,
it is not practical to consider a ratio of nuclear to nucleon
structure functions as in the unpolarized case in Fig.~\ref{fig1}.)
A larger effect arises with the addition of $\Delta$ contributions to the
$^3$He wave function \cite{BGST02}, which accentuates the differences
between the free neutron and $^3$He structure functions, especially
at intermediate values of $x$, $0.1 \lesssim x \lesssim 0.4$.
Qualitatively similar behavior is seen for the $g_2$ structure function
in Fig.~\ref{fig2}(d), with the signs reversed compared to $g_1$.
Namely, the neutron polarization slightly reduces the (positive)
$g_2^n$ contribution, while the (negative) proton polarization
combines with the (negative) $g_2^p$ structure function to produce
a compensating shift upward, leaving $g_2^{^3{\rm He}} > g_2^n$.
Again, this trend is also seen in the DIS part of the $g_2$
comparison at $Q^2=1$~GeV$^2$ in Fig.~\ref{fig2}(b).

\begin{figure}
\includegraphics[width=10cm]{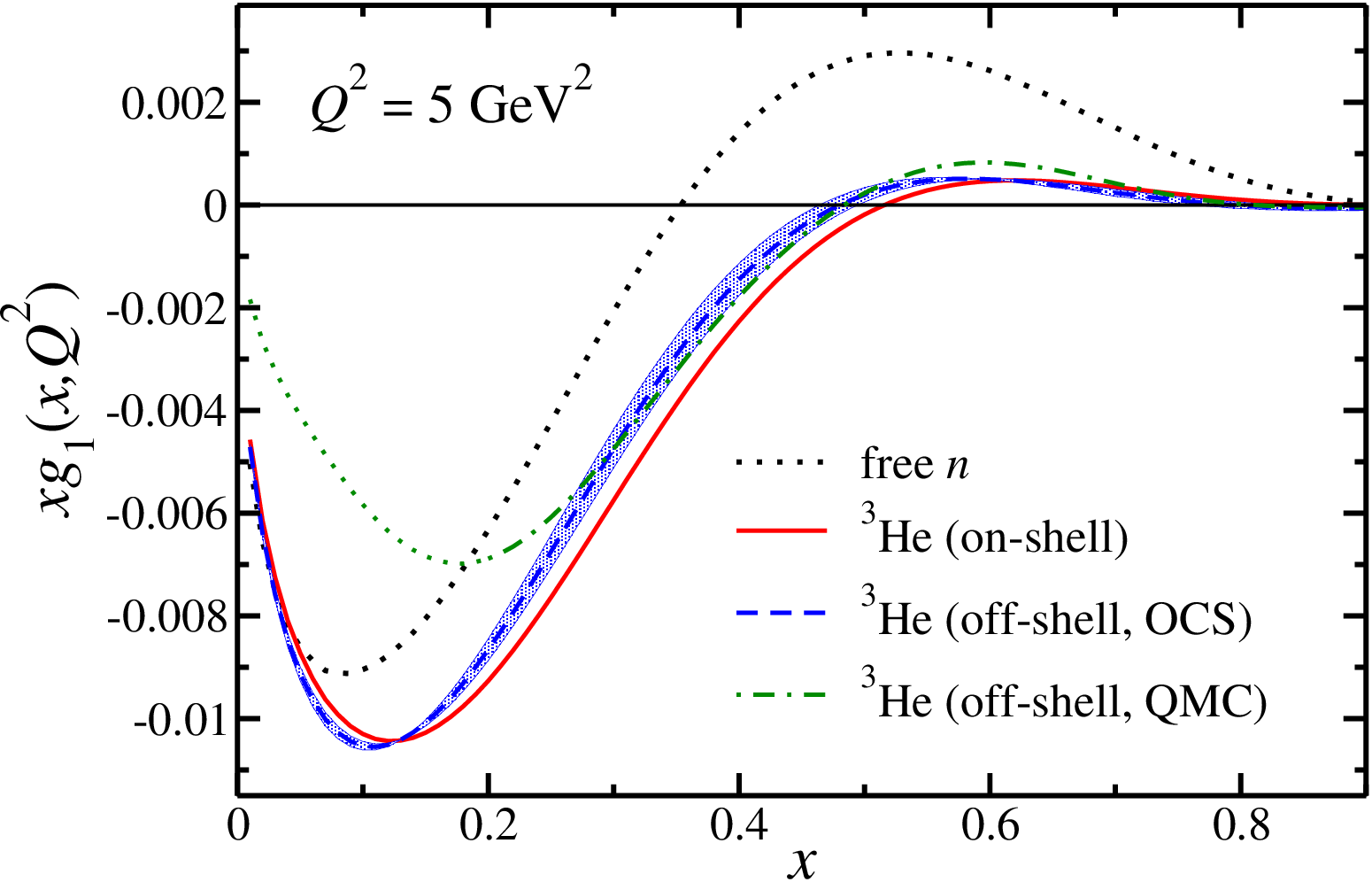}
\caption{Nucleon off-shell corrections to the $xg_1$ structure function
	of $^3$He, within the off-shell covariant spectator model
	(dashed and shaded band) and in the quark-meson coupling model
	\cite{Steffens99} (dot-dashed) in the valence approximation,
	with an extrapolation of the latter for $x \lesssim 0.2$
	(light dot-dashed).
	The free neutron structure function (dotted) and the $^3$He
	structure function computed with on-shell nucleon input
	(solid) are shown for comparison.}
\label{fig3}
\end{figure}

The effects of the (negative) $\Delta$ resonance contribution to
$g_1^{^3{\rm He}}$ are offset somewhat by the nucleon off-shell
corrections discussed in Sec.~\ref{ssec:off}.  As illustrated in
Fig.~\ref{fig3}, the corrections computed within the OCS model give
rise to a positive contribution in the intermediate-$x$ region,
$0.1 \lesssim x \lesssim 0.6$, where the magnitude of the effects
is largest.  This will mostly cancel the impact of the $\Delta$
resonance in this region, bringing the total $^3$He structure
function closer to the on-shell result.  (For simplicity, here we
have computed the effects of the smearing in the $\gamma=1$ limit,
although as Fig.~\ref{fig2} illustrates, at $Q^2 = 5$~GeV$^2$
the finite-$Q^2$ effects are negligible.)
To give an estimate of the uncertainty on this correction,
we consider a range of nucleon swelling parameters $\delta R_N$
between $\delta R_N/R_N \approx 2\%$ and 6\%, with a central
value of 4\%.  This gives for the parameter $\lambda_N$ which
determines the $p^2$ derivative $\partial g_i^N/\partial p^2$
in Eqs.~(\ref{eq:dqdp2})--(\ref{eq:lamndaN}) the values
$\lambda_n = 0.84 \pm 0.42$ for the neutron and
$\lambda_n = 1.12 \pm 0.56$ for the proton.
The corrections corresponding to this range of parameters
is indicated in Fig.~\ref{fig3} by the shaded band.

Qualitatively similar behavior is observed using the QMC off-shell
model from Steffens {\it et al.} \cite{Steffens99}, which gives a
small positive shift in $g_1^{^3{\rm He}}$ over most of the $x$
range considered.  Note that this model assumes the valence quark
approximation, so that its predictions at small $x$ ($x \lesssim 0.2$)
may not be reliable.  Nevertheless, it is reassuring that these
models, which are based on rather different assumptions, lead to
off-shell corrections that are similar in sign and magnitude in
their regions of validity.

\begin{figure}
\includegraphics[width=8cm]{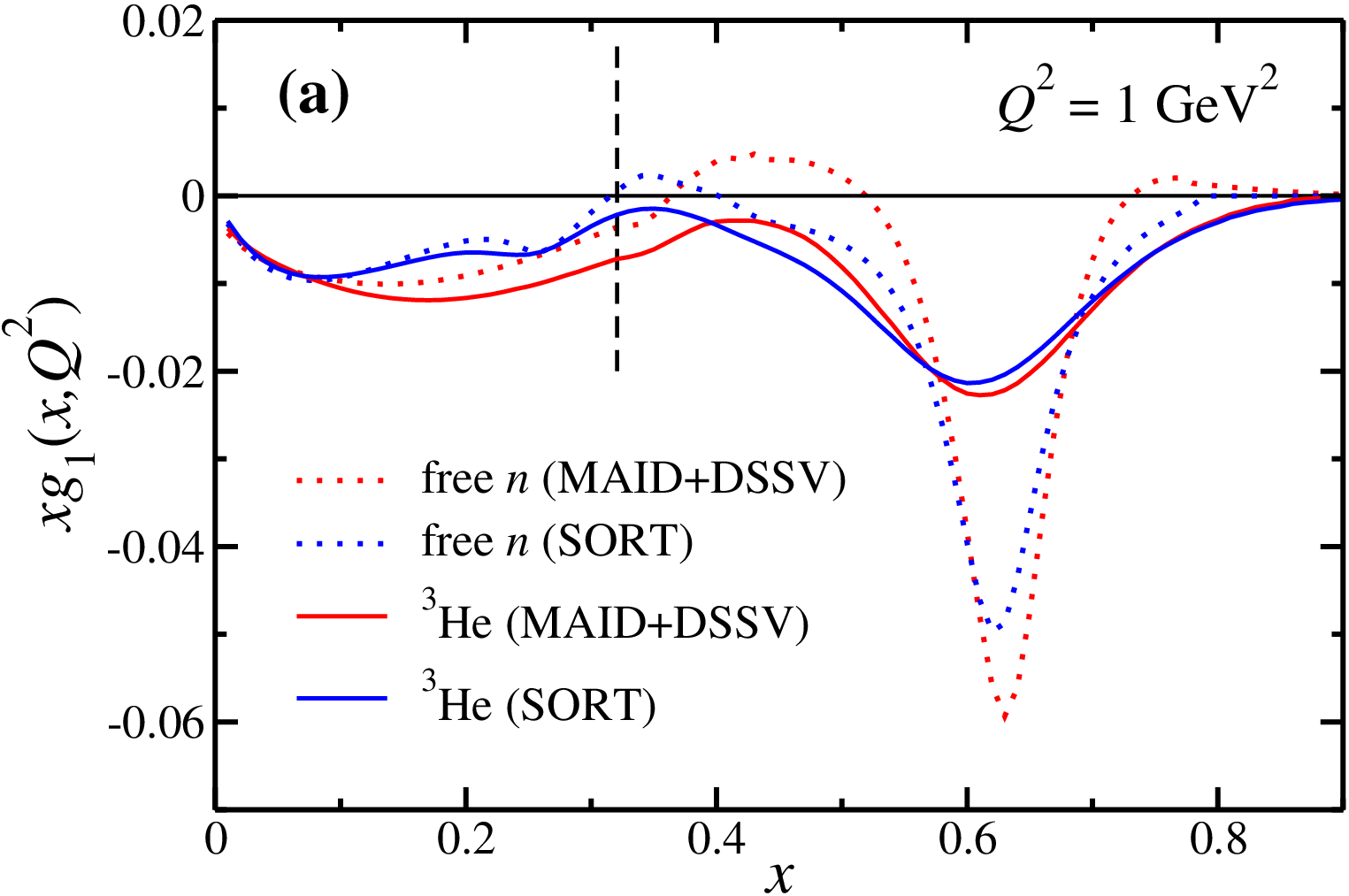}
\includegraphics[width=8cm]{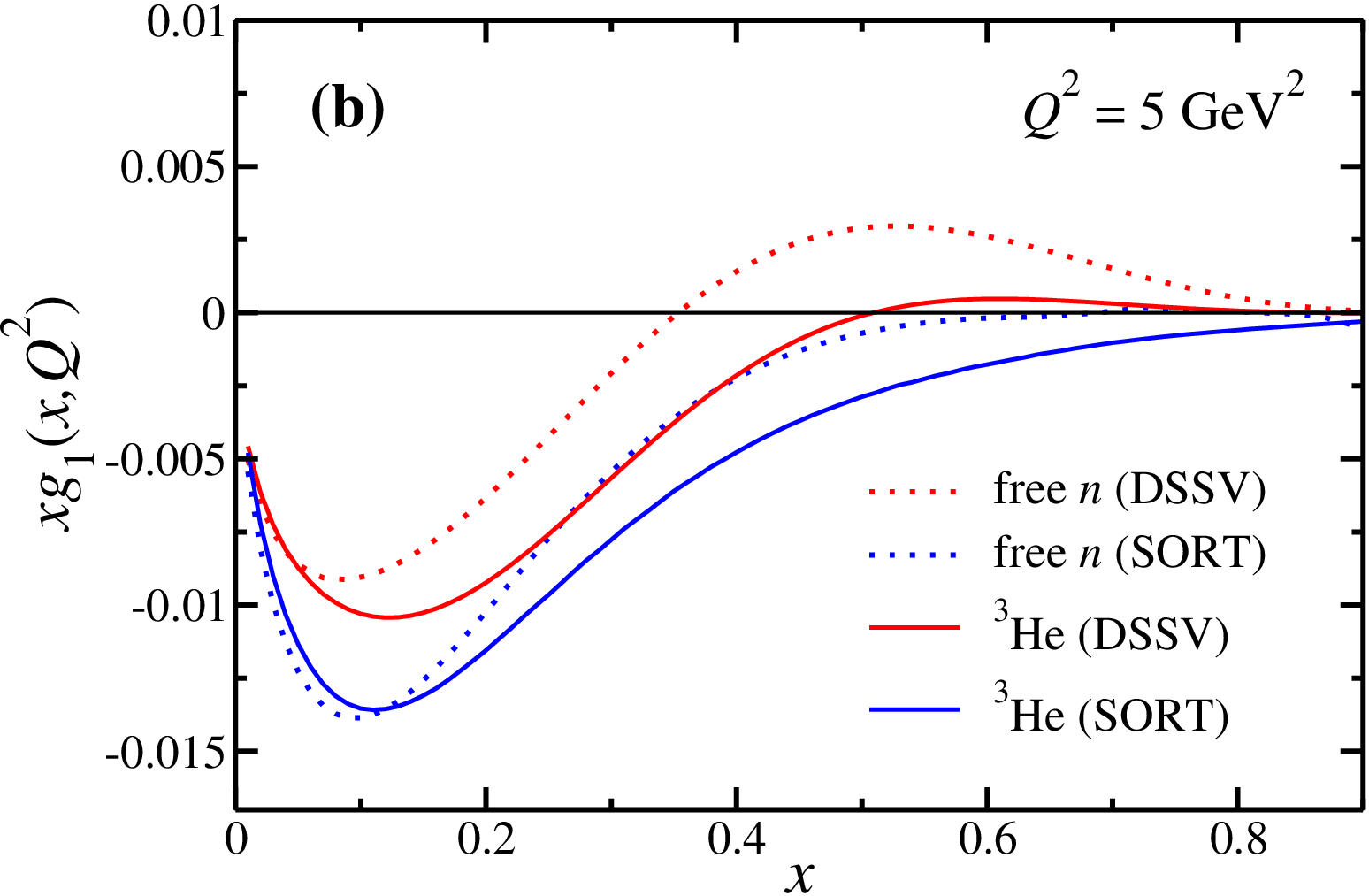}\\
\vspace*{0.9cm}
\includegraphics[width=8cm]{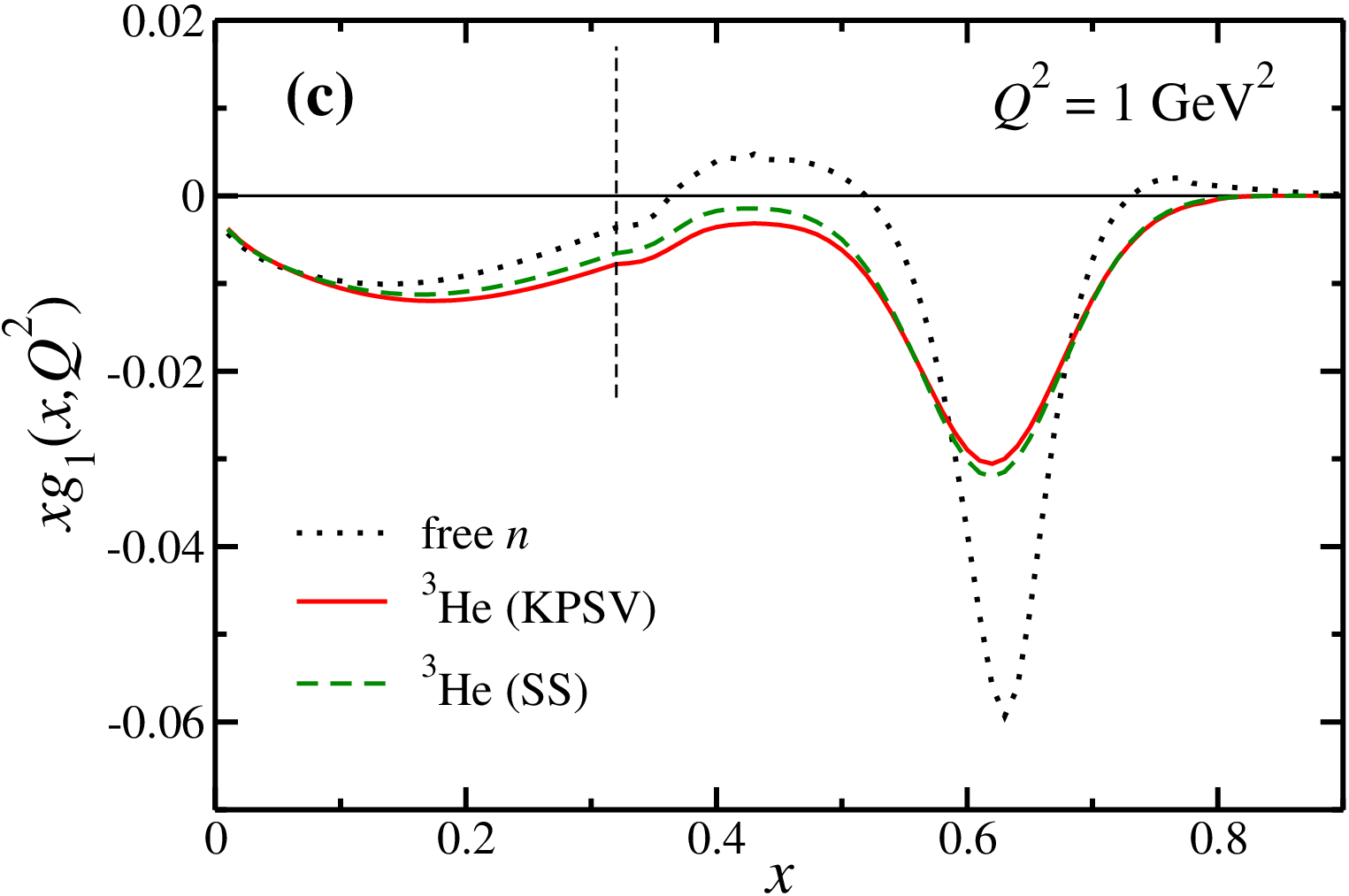}
\includegraphics[width=8cm]{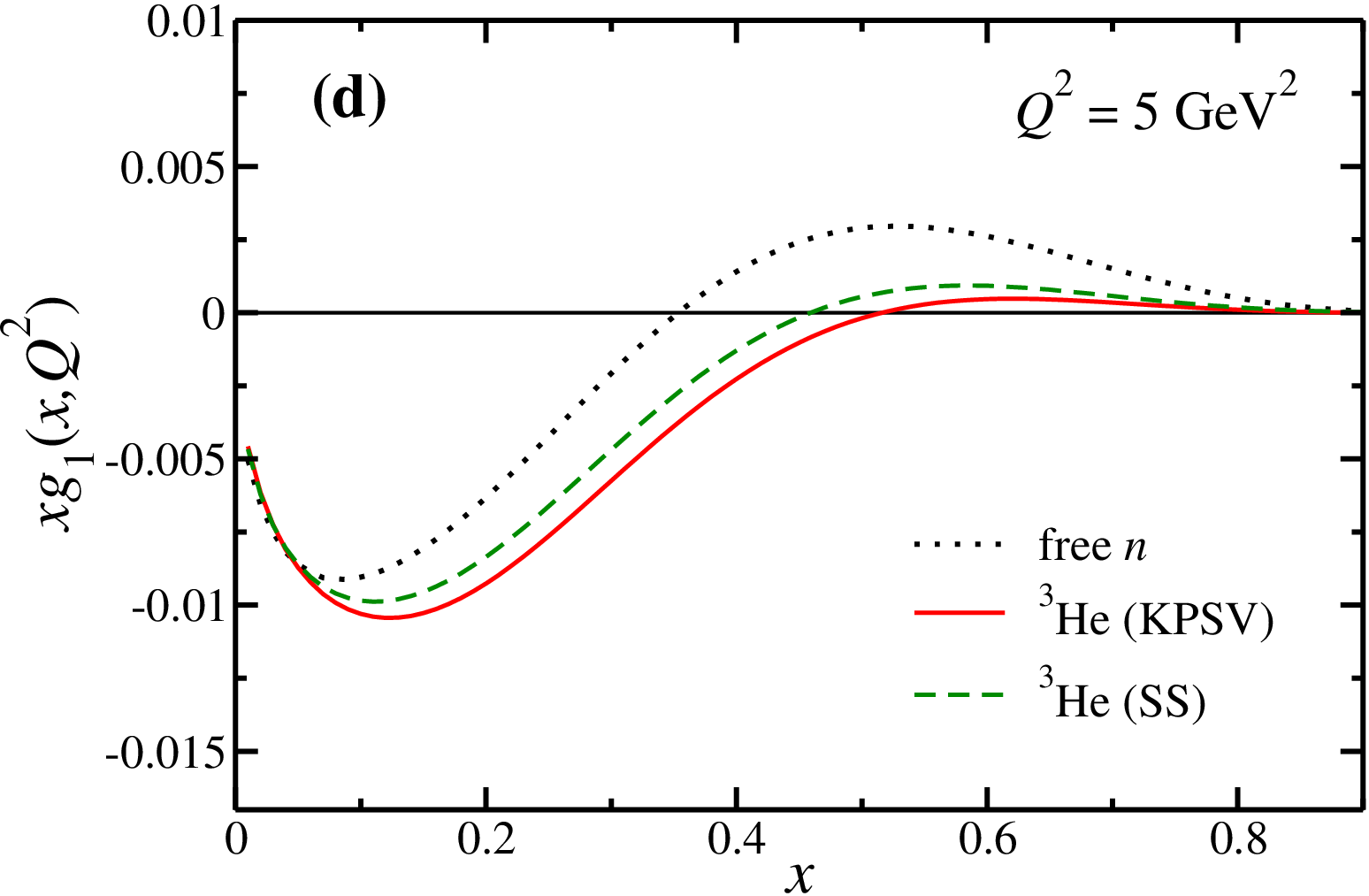}
\caption{Dependence of the neutron and $^3$He $g_1$ structure
	functions on the input nucleon parametrization (MAID
	\cite{MAID99} and DSSV \cite{DSSV09}, and SORT
	\cite{Simula02}) [{\bf (a)} and {\bf (b)}],
	and on the $^3$He wave function (KPSV \cite{KPSV}
	and SS \cite{SS}) [{\bf (c)} and {\bf (d)}],
	at $Q^2 = 1$~GeV$^2$ and 5~GeV$^2$.
	The dashed vertical lines at $Q^2 = 1$~GeV$^2$ indicate the
	boundary between the nucleon resonance and DIS regions.}
\label{fig4}
\end{figure}

While the above results are obtained using specific parametrizations
for the input proton and neutron $g_1$ and $g_2$ structure functions
\cite{MAID99, DSSV09}, the detailed predictions for the $^3$He
functions will naturally be modified with different nucleon inputs.
In Figs.~\ref{fig4}(a) and (b) we compare the results for the neutron
and $^3$He structure functions, computed using the full smearing
functions in Eq.~(\ref{eq:2Dconv}), as in Fig.~\ref{fig2}, with those
using the combined parametrization of the resonance and DIS regions
by Simula {\it et al.} (SORT) \cite{Simula02}.  The quantitative
differences between the two sets of results reflect the degree to
which the structure functions of the free nucleon, and particularly
the neutron, are determined experimentally.  However, the conclusions
about the relative importance of the various nuclear corrections
investigated here does not depend on the form of the input distributions.
Furthermore, the dependence of the $^3$He structure functions on the
input neutron $g_1$ and $g_2$, can in principle be removed by applying
an iterative procedure, such as that outlined by Kahn {\it et al.}
\cite{Kahn09}, for example.  In practice, the number of iterations
required for convergence depends on the number of data points and
the precision of the data.

Indeed, the only theoretical input on which the $^3$He structure
functions in principle depend are the nuclear smearing functions,
and whatever approximations are made for them.
In Figs.~\ref{fig4}(c) and (d) the dependence on the nuclear
structure model is illustrated by comparing the $g_1$ structure
function computed from the KPSV \cite{KPSV} and SS \cite{SS}
$^3$He wave functions, at $Q^2 = 1$~GeV$^2$ and 5~GeV$^2$.
To isolate the effects of the wave function alone, the same Bjorken
limit ($\gamma=1$) approximation is used for the smearing functions
in the two models.  The results in both the resonance region and in
the DIS region show a very mild dependence on the wave function,
smaller than on the input nucleon structure functions in 
Figs.~\ref{fig4}(a) and (b), suggesting that the theoretical
uncertainty arising from the nuclear wave function is not significant.

\begin{figure}[tb]
\includegraphics[width=8cm]{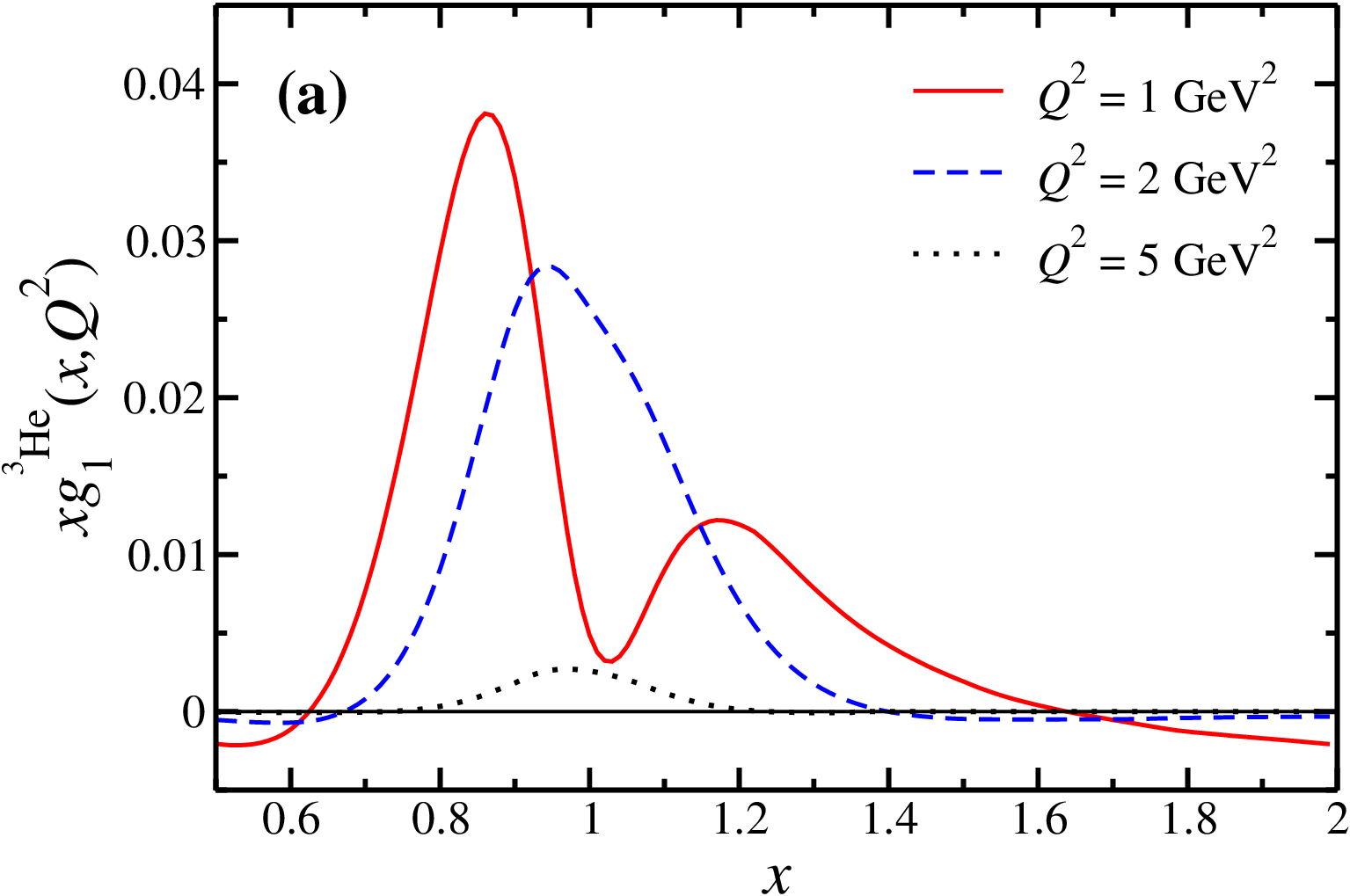}
\includegraphics[width=8cm]{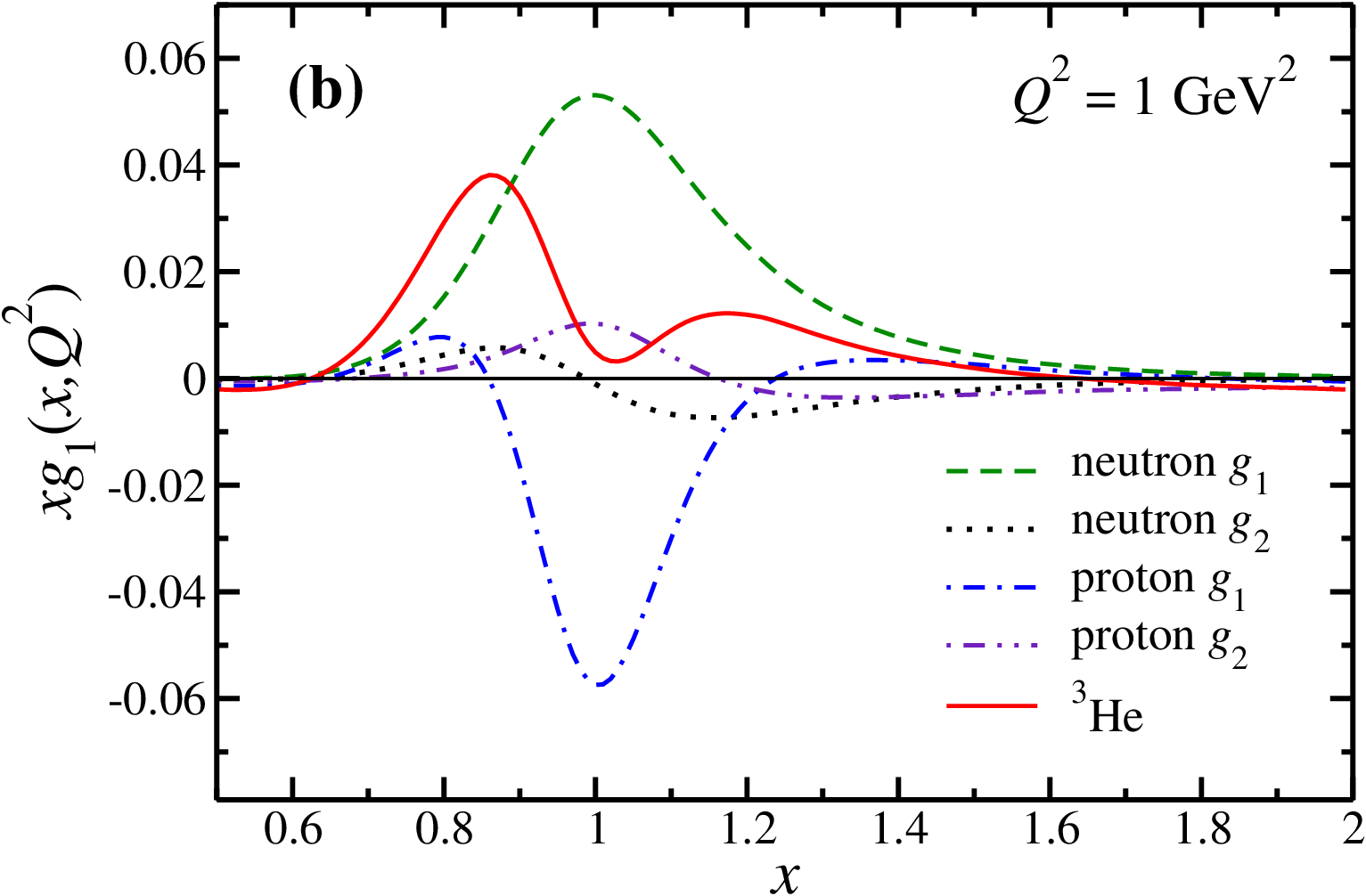}
\caption{{\bf (a)}
	Quasi-elastic contributions to the $xg_1$ structure function of
	$^3$He at $Q^2 = 1$ (solid), 2 (dashed) and 5~GeV$^2$ (dotted).
	 {\bf (b)}
	Individual contributions to the quasi-elastic $xg_1$ structure
	function of $^3$He (solid) at $Q^2 = 1$~GeV$^2$, from the
	neutron $g_1$ (dashed), neutron $g_2$ (dotted),
	proton $g_1$ (dot-dashed) and proton $g_2$ (dot-dot-dashed)
	structure functions.}
\label{fig5}
\end{figure}

The dependence on the $^3$He wave function model can be reduced by
comparing the calculated smearing functions with data on quasi-elastic
cross sections.  In the impulse approximation these depend simply on the
product of the smearing function and the nucleon elastic form factors,
Eqs.~(\ref{eq:QE}).  To the extent that the form factors are determined
from data in other elastic or QE scattering reactions, and corrections
from final state interactions or non-nucleonic contributions are not
large, measurement of the QE cross sections can directly constrain the
smearing function in the vicinity of the QE peak.

The predictions for the QE contribution to the $g_1$ structure function
of $^3$He are shown in the WBA in Fig.~\ref{fig5}(a) for $Q^2 = 1$, 2
and 5~GeV$^2$, using the full, $Q^2$-dependent smearing functions in
Eqs.~(\ref{eq:Dy}) and (\ref{eq:Dij}).  For the proton electric and
magnetic elastic form factors we use the parametrization from
Ref.~\cite{AMT07}, while the Kelly \cite{Kelly04} fit is used for
the neutron form factors.
As expected, the amplitude of the QE peak fall rapidly with increasing
$Q^2$, so that by $Q^2 = 5$~GeV$^2$ the QE contribution is strongly
suppressed.  At the lowest $Q^2$ value, $Q^2 = 1$~GeV$^2$, the QE
contribution exhibits a striking double peak structure, with local
maxima at $x \approx 0.85$ and 1.2, and a dip at $x \approx 1$.
This is due primarily to the proton $g_1^p$ contribution, which has
a relatively large and negative peak at $x=1$, as Fig.~\ref{fig5}(b)
illustrates.  Although the polarized proton smearing function in $^3$He
is strongly suppressed relative to the neutron, the larger proton
electric form factor $G_E^p$ compared with the neutron $G_E^n$ makes
the proton and neutron contributions comparable at this $Q^2$ value.

Future data on inclusive QE cross sections in the $x \sim 1$ would
allow one to investigate this intriguing interplay between the various
components of $g_1^{^3{\rm He}}$ in detail, and provide a sensitive
test of the nuclear wave function.  Examination of the tails of the
QE cross sections at large $x$ ($x \gtrsim 1.6$) would also enable
exploration of, and possible constraints on, the wave function and
nucleon off-shell effects \cite{QE13}.

% .......................................................................
\subsection{Asymmetries}
\label{ssec:asym}

While nuclear effects in structure functions have received the greatest
attention theoretically, the quantities that are most directly accessible
in polarized DIS experiments are the polarization asymmetries $A_1$ and
$A_2$ in Eqs.~(\ref{eq:asym}) (which are themselves extracted from the
longitudinal and parallel asymmetries in Eqs.~(\ref{eq:Apar12})).
As ratios of spin-dependent to spin-averaged structure functions,
the polarization asymmetries can display more subtle effects arising
from the $x$ dependence of the polarized $g_{1,2}$ and unpolarized
$F_{1,2}$ structure functions, and the nuclear corrections to these,
especially in the resonance nucleon region.

\begin{figure}[t]
\includegraphics[width=8cm]{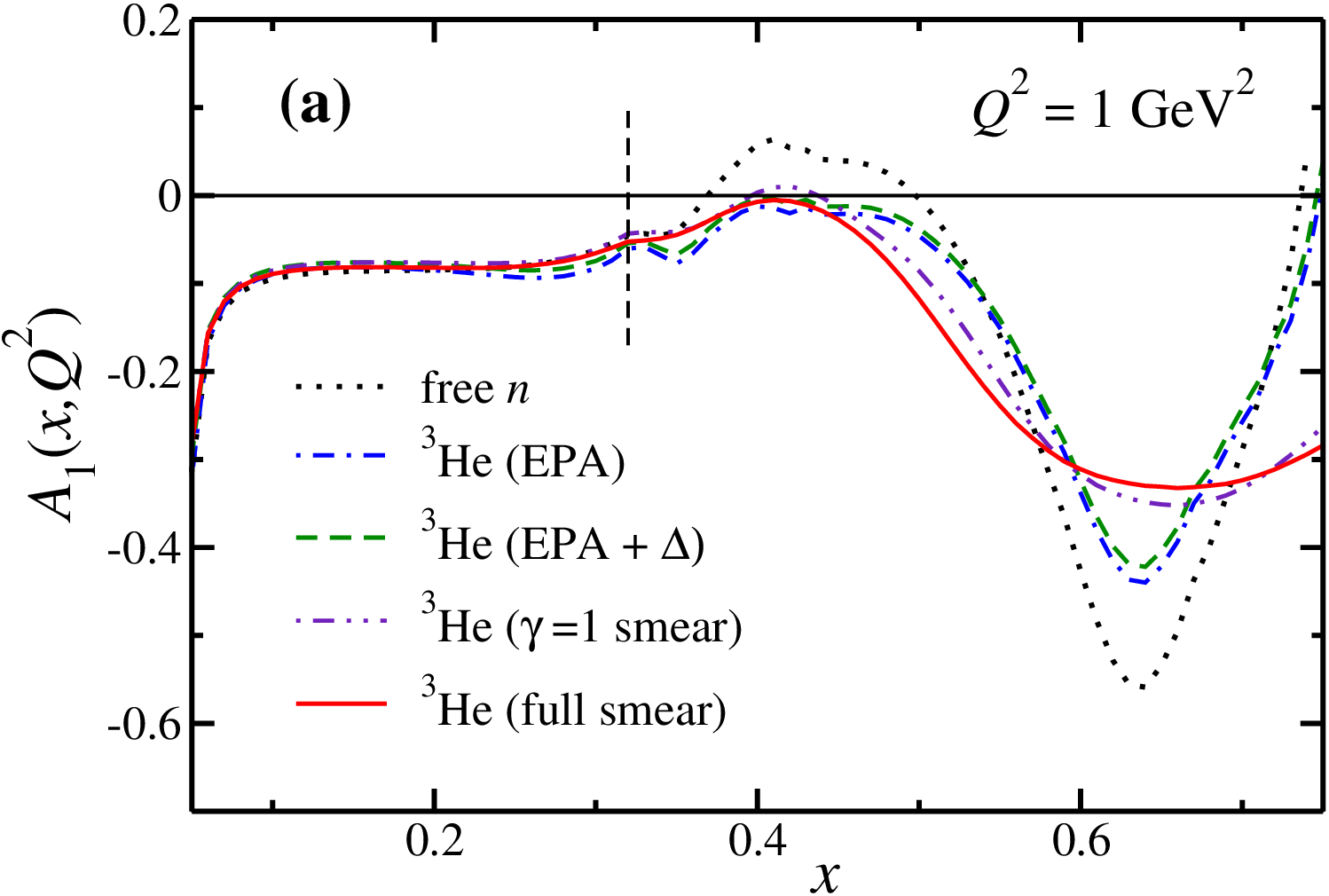}
\includegraphics[width=8cm]{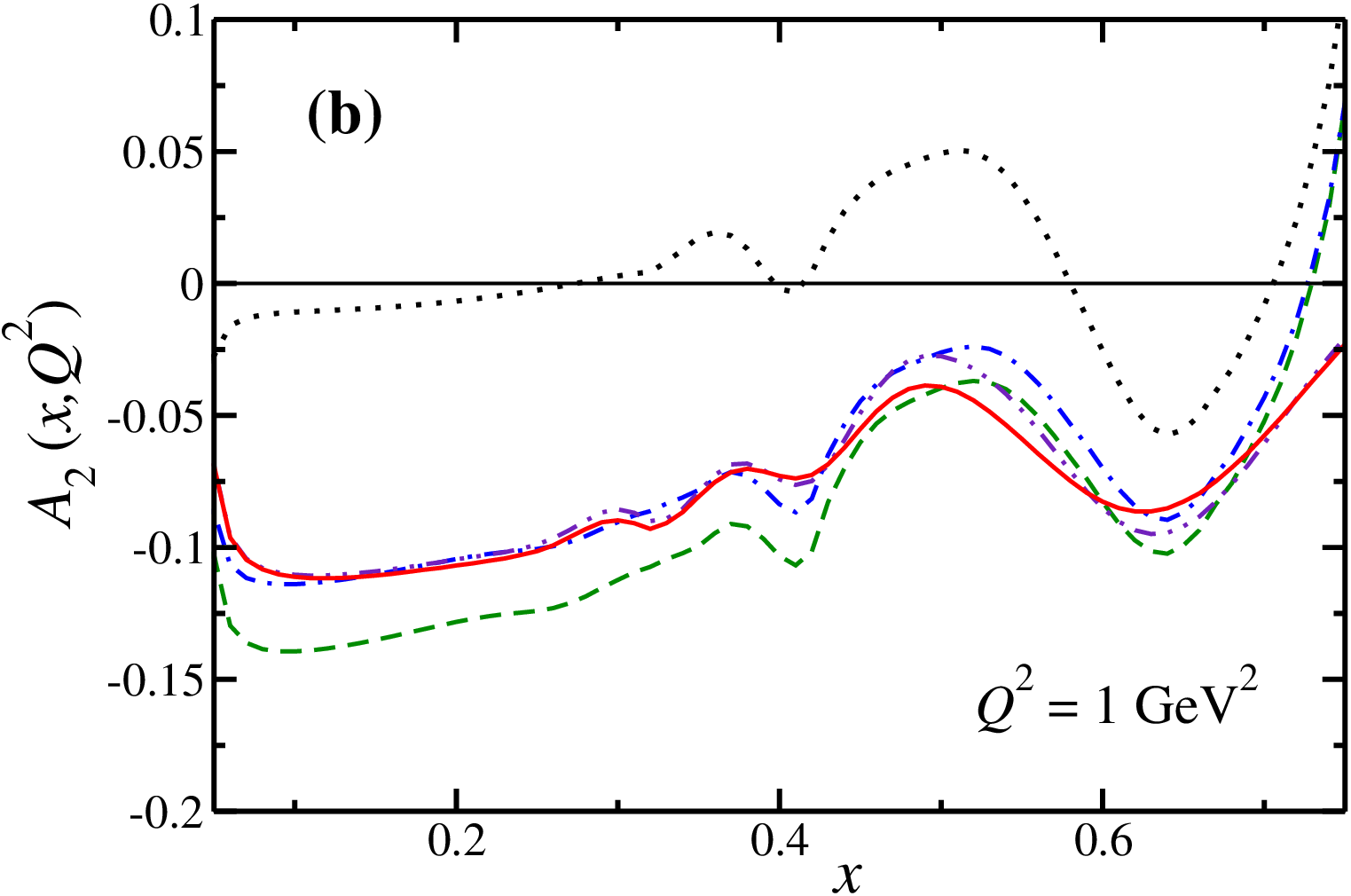}\\
\vspace*{0.9cm}
\includegraphics[width=8cm]{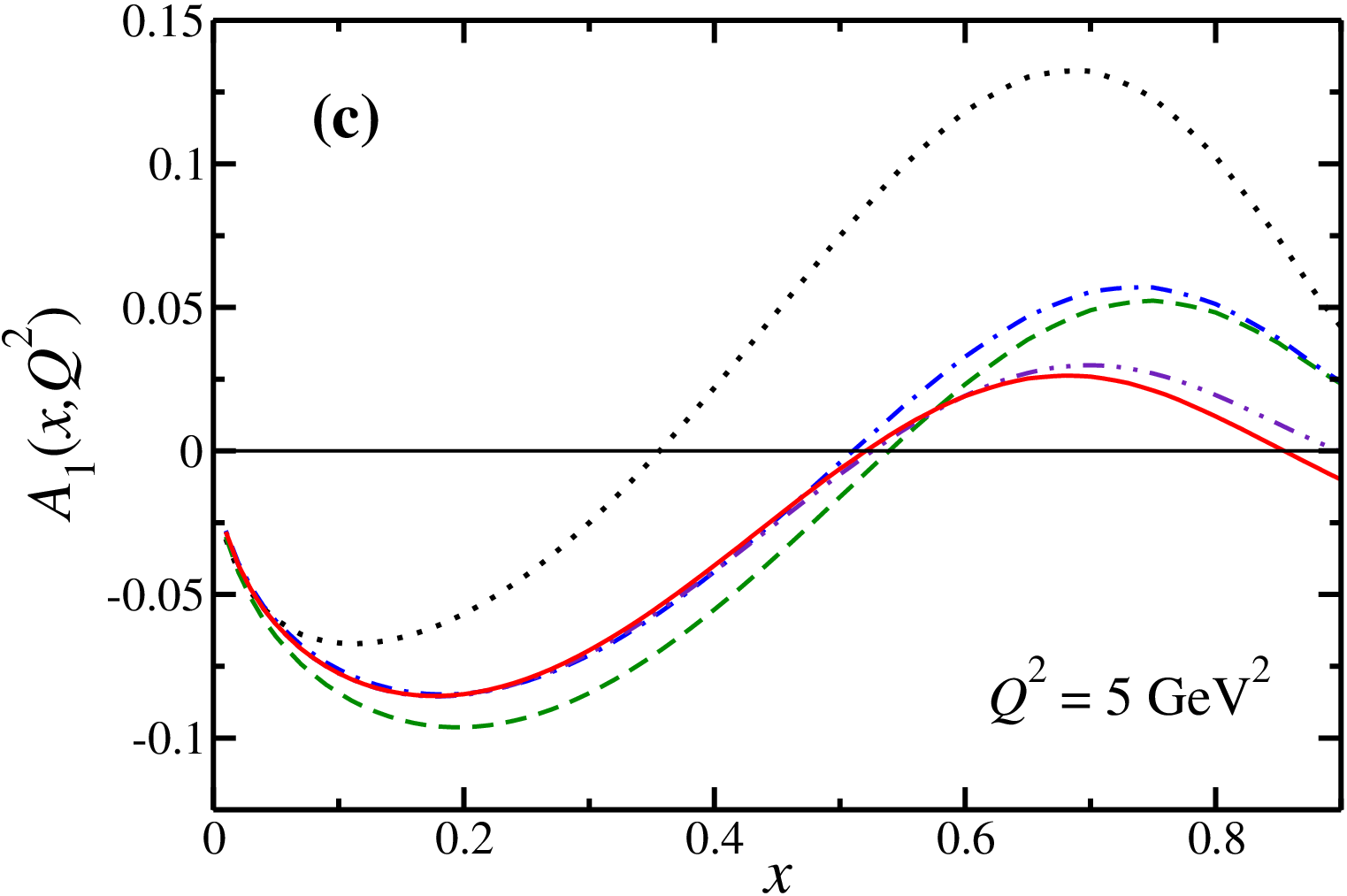}
\includegraphics[width=8cm]{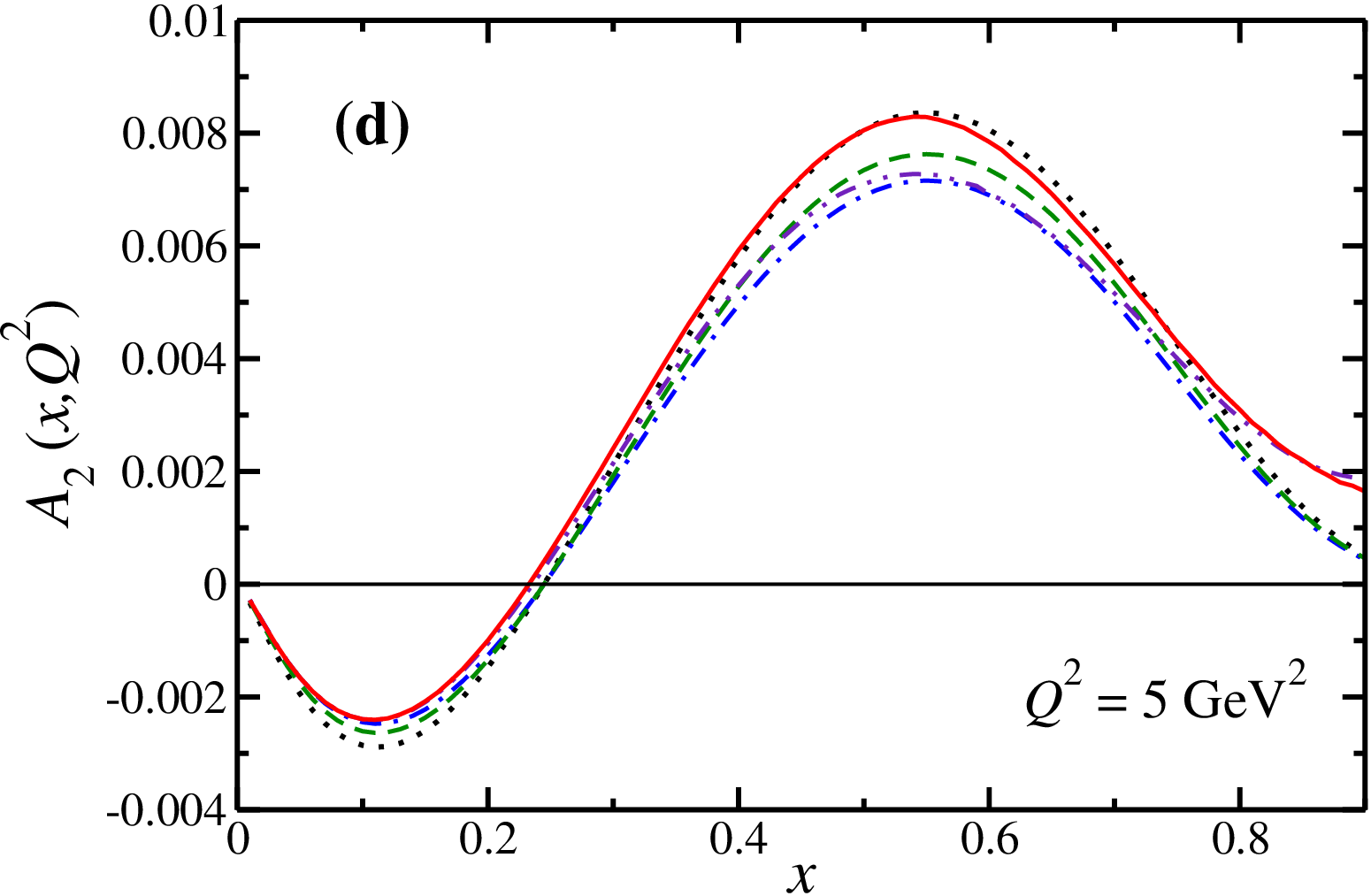}
\caption{As in Fig.~\ref{fig2}, but for the polarization
	asymmetries $A_1$ and $A_2$ of the neutron and $^3$He at
	$Q^2 = 1$~GeV$^2$ [{\bf (a)} and {\bf (b)}] and
	$Q^2 = 5$~GeV$^2$ [{\bf (c)} and {\bf (d)}],
	constructed from ratios of the spin-dependent structure 
	functions in Fig.~\ref{fig2} and the unpolarized $F_1$
	structure function from the Bosted-Christy parametrization
	\cite{BC}.  Note that the $^3$He asymmetries are scaled by
	a factor $(1 + 2F_1^p/F_1^n)$.}
\label{fig6}
\end{figure}

In Fig.~\ref{fig6} the $A_1$ and $A_2$ asymmetries of the neutron
and $^3$He are shown at $Q^2 = 1$ and 5~GeV$^2$ for the various
nuclear models considered in Fig.~\ref{fig2}.  Note that since
$g_{1,2}^{^3{\rm He}} \approx g_{1,2}^n$, while
$F_1^{^3{\rm He}} \gg F_1^n$, the absolute value of the $^3$He asymmetry
will be considerably smaller than that of the neutron asymmetry.
To display both the neutron and $^3$He asymmetry results on the same
scale, we multiply the latter by the factor $(1 + 2 F_1^p / F_1^n)$,
which compensates for the suppression of $A_{1,2}^{^3{\rm He}}$ due to
the small proton contribution to $g_{1,2}^{^3{\rm He}}$,
\begin{subequations}
\label{eq:He3asym}
\begin{eqnarray}
A_1^{^3{\rm He}}
&=& {\left( g_1^{^3{\rm He}} - (\gamma^2-1) g_2^{^3{\rm He}} \right)
     \over F_1^{^3{\rm He}}}\ \
\longrightarrow\
A_1^{^3{\rm He}} 
\times \left( 1 + {2 F_1^p \over F_1^n} \right),	\\
A_2^{^3{\rm He}} 
&=& \sqrt{\gamma^2-1}\,
    {\left( g_1^{^3{\rm He}} + g_2^{^3{\rm He}} \right)
     \over F_1^{^3{\rm He}}}\
\longrightarrow\
A_2^{^3{\rm He}}
\times \left( 1 + {2 F_1^p \over F_1^n} \right).
\end{eqnarray}
\end{subequations}

For the $A_1$ asymmetry, the effect of the nuclear corrections is
qualitatively similar to that for the $g_1$ structure function in
Fig.~\ref{fig2}.  The nuclear smearing corrections have the largest
impact in the nucleon resonance region, particularly in the vicinity
of the $\Delta$ resonance, although the magnitude of the effect is
slightly smaller compared to that for $g_1$.  Since both the numerator
and denominator in the $^3$He asymmetry involve smeared structure
functions, the relative effects of the smearing on $A_1^{^3{\rm He}}$
will be reduced.  Note that for the isovector $\Delta$ resonance,
the scaling factor in Eqs.~(\ref{eq:He3asym}) is
$1 + 2F_1^p/F_1^n \approx 3$.
At larger $x$, in the region $W \lesssim M_\Delta$, the free neutron
asymmetry computed from the MAID parametrization of $g_1$ and $g_2$
rises steeply, which suggests that the $F_1^n$ denominator from the
Bosted-Christy fit falls rapidly in this region.  A similar trend is
observed when using the SORT parametrization \cite{Simula02} of the
spin-dependent structure functions.
The general features of the $A_1$ asymmetry in the DIS region at
$Q^2=5$~GeV$^2$ are similar to those of the $g_1$ structure function
in Fig.~\ref{fig2}(c).  Namely, with the rescaling factor in
Eqs.~(\ref{eq:He3asym}), the asymmetries at $x \ll 1$ are basically
given by the corresponding $g_1$ structure functions divided by the
$F_1^n$ structure function, with the scaled $^3$He asymmetry below
the free neutron asymmetry.

For the $A_2$ asymmetry, which is proportional to the sum of the
$g_1$ and $g_2$ structure functions, the resulting scaled $^3$He
asymmetry at $Q^2=5$~GeV$^2$ is very similar to the input $A_2^n$.
Here, the dominant leading twist contribution to $A_2$ is given by
the integral term $\int (dz/z) g_1(z)$ on the right-hand-side of
Eq.~(\ref{eq:WW}).  Therefore, the differences between the $A_2$
asymmetries of the neutron and $^3$He at leading twist will be
determined by the nuclear effects on the $g_1$ structure function.
As is evident from Figs.~\ref{fig2}(c) and (d), the nuclear
corrections lower the $^3$He structure function relative to the
neutron for $g_1$ but raise it for $g_2$, the net effect of which
is a strong cancellation of the nuclear effects (for both the EPA
and smearing calculations) which leaves
$A_2^n \approx A_2^{^3{\rm He}} (1+2F_1^p/F_1^n)$
at $x \ll 1$.
The cancellations are not as evident at the lower, $Q^2=1$~GeV$^2$
value, where the resonance structures dominate and the WW approximation
(\ref{eq:WW}) to $g_2$ is in general not valid.  Here the prominent
$\Delta$ resonance peaks in $g_1$ (negative) and $g_2$ (positive) in
Fig.~\ref{fig2} largely cancel, resulting in a $^3$He $A_2$ asymmetry
that is several times smaller than the corresponding $A_1$ asymmetry
at the $\Delta$ peak in Fig.~\ref{fig6}(a).
Because of the $\sqrt{\gamma^2-1}$ factor in the definition of
$A_2$ in Eq.~(\ref{eq:A2}), the overall magnitude of the $A_2$
asymmetry at lower $Q^2$ (larger $\gamma$) is almost an order of
magnitude larger than that at $Q^2=5$~GeV$^2$ in Fig.~\ref{fig6}(d).

\begin{figure}[t]
\includegraphics[width=8cm]{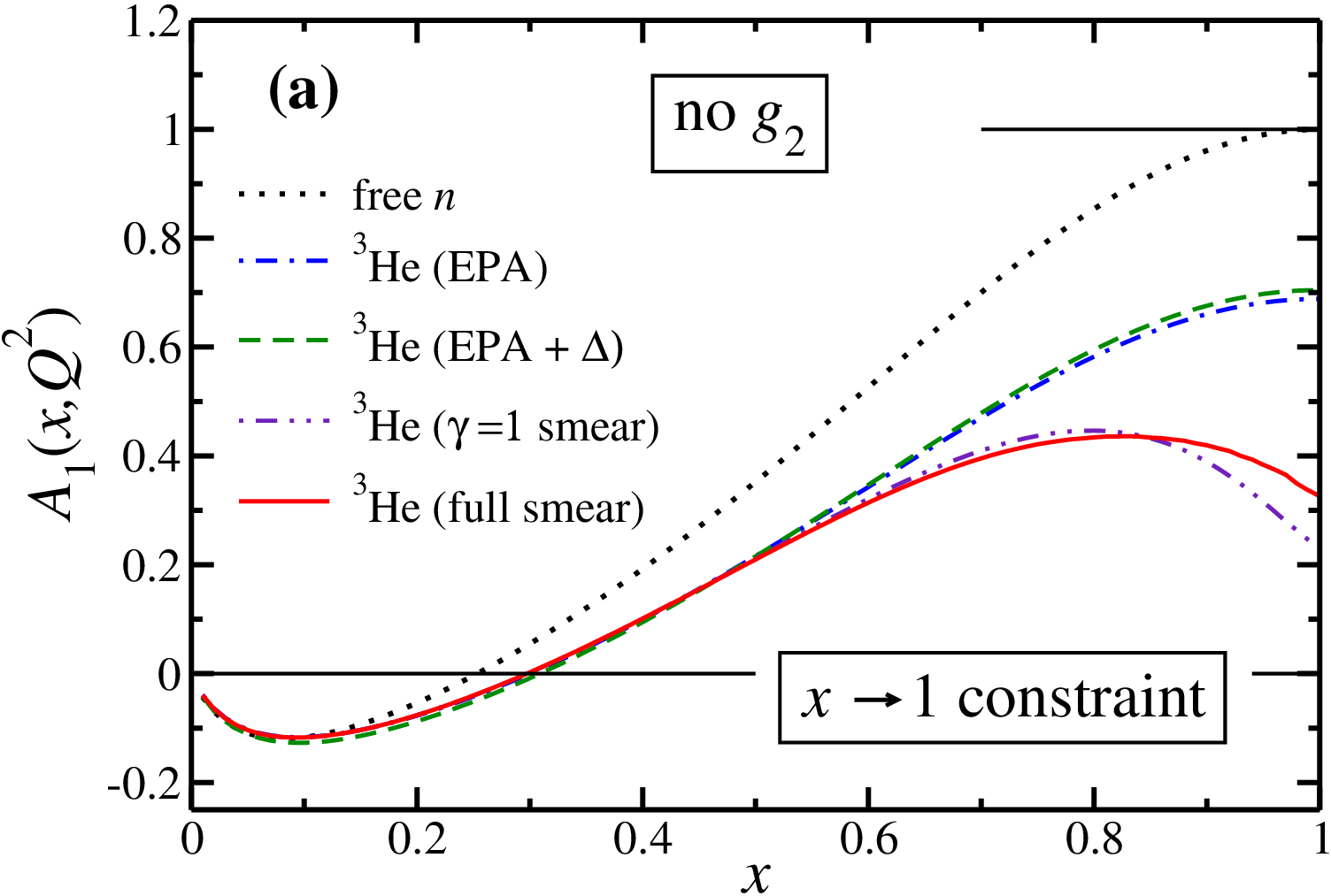}
\includegraphics[width=8cm]{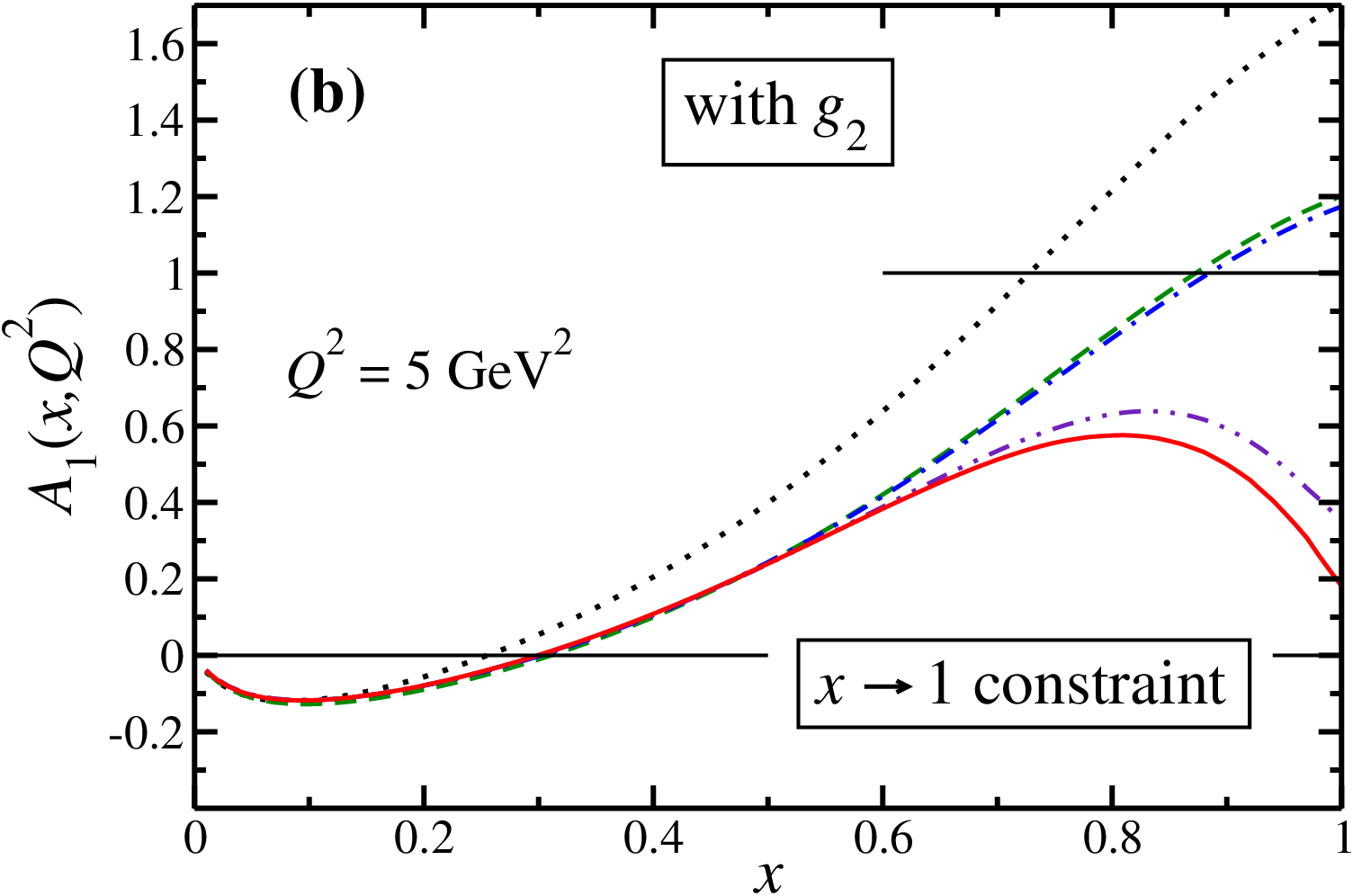}
\caption{Polarization asymmetry $A_1$ of the neutron and $^3$He
	computed using the input LSS \cite{LSS98} parametrization
	constrained by $A_1 \to 1$ as $x \to 1$.
	The asymmetries are computed
	{\bf (a)} with $g_1$ contributions only and
	{\bf (b)} including also $g_2$ corrections in
	Eq.~(\ref{eq:A1}), which push the $A_1$ asymmetry
	above unity at large $x$ at $Q^2=5$~GeV$^2$.
	The nuclear models are as in Fig.~\ref{fig6},
	and the the $^3$He asymmetry is scaled by a factor
	$(1 + 2 F_1^p/F_1^n)$.}
\label{fig7}
\end{figure}

From simple counting rule and perturbative QCD arguments, in the
$x \to 1$ limit DIS from quarks with spins aligned with the spin 
of the nucleon is expected to dominate over scattering from quarks
with spins antialigned \cite{Farrar75, Blankenbecler74, Gunion73,
BrodskyLepage79, BBS95}.  At leading twist, the proton and neutron
polarization asymmetries should therefore approach unity, $A_1 \to 1$
as $x \to 1$.  A number of other, nonperturbative models also make  
specific predictions for the large-$x$ behavior of $A_1^n$, making
this quantity particularly sensitive to the dynamics of valence
quarks in the nucleon.
Because of the lack of data on spin structure functions or asymmetries
at very large $x$, however, the $x \to 1$ behavior is usually not
addressed in standard PDF parametrizations, such as the DSSV fit 
\cite{DSSV09} used in Fig.~\ref{fig6}, and the behavior at
$x \gtrsim 0.8$ is left unconstrained.
To illustrate the possible effects of the perturbative $x \to 1$
expectations on the spin-dependent PDFs, Leader, Sidorov and Stamenov
\cite{LSS98} performed a global fit with polarized and unpolarized
PDFs constrained with $\Delta q/q \to 1$ as $x \to 1$ \cite{BBS95},
which at large $Q^2$ forces $A_1 \to g_1/F_1 \to 1$.
The neutron asymmetry with this constraint is illustrated in
Fig.~\ref{fig7}(a), where for simplicity the $g_2$ contribution
is omitted.

The question we would like to address in this work is how the
nuclear corrections in $^3$He would affect such behavior, and
the degree to which these corrections can be reliably subtracted
to reveal the true dependence of $A_1^n$ on $x$ as $x \to 1$.
Within the EPA, the $^3$He asymmetry is reduced by $\approx 30\%$
in the $x \to 1$ limit relative to the neutron asymmetry,
for the same reasons that the effective polarizations render
$g_1^{^3{\rm He}} < g_1^n$ in Fig.~\ref{fig2}(c), for example,
and that the scaled $^3$He asymmetry lies below $A_1^n$ in
Fig.~\ref{fig6}(c) for the DSSV parametrization.
The effect of the nuclear smearing is a further reduction of    
$A_1^{^3{\rm He}}$ to $\approx 0.25 - 0.35$ at $Q^2 = 5$~GeV$^2$,
depending on whether the full, $Q^2$-dependent smearing function  
is used or its $\gamma=1$ approximation.
Including the $g_2$ terms in Eq.~(\ref{eq:A1}), the free neutron     
$A_1$ asymmetry, computed from the LSS parametrization \cite{LSS98} 
using the WW relation (\ref{eq:WW}) for $g_2^n$, increases by     
$\gtrsim 60\%$ at $x \approx 1$ for $Q^2=5$~GeV$^2$, as
Fig.~\ref{fig7}(b) illustrates.  (Recall from Fig.~\ref{fig2}(d)
that the twist-2 part of $g_2^n$ is negative at large $x$.)
The resulting $^3$He asymmetries are correspondingly larger,
although the effects of the $Q^2$-dependent smearing are even    
more pronounced in the presence of the $g_2$ contributions.      
With the upcoming high-precision experiments to determine the
$x \to 1$ behavior of $A_1^n$ from measurements of the $^3$He   
polarization asymmetries planned at Jefferson Lab at 12~GeV
\cite{PR12-06-110, PR12-06-122}, it will therefore be crucial to  
account for the finite-$Q^2$ and nuclear smearing corrections
in the large-$x$ region.

% .......................................................................
\subsection{Moments}
\label{ssec:mom}

The nuclear corrections examined in this analysis clearly have
a significant impact on the shape of the structure functions,
especially at large values of $x$.
Since in this region the structure functions themselves are typically
small, one may expect that the nuclear effects on integrals of
structure functions, or moments $\Gamma^{(n)}_i$, may be reduced.
This would be expected particularly of the low moments, which are most
sensitive to the small-$x$ region, whereas higher moments progressively
emphasize the large-$x$ tails of distributions with increasing rank $n$.

In QCD, the moments of structure functions are formally related
through the operator product expansion to hadronic matrix elements of
local operators of a given twist, and can be directly computed from
first principles in lattice QCD or approximated in low-energy model 
calculations.  Various sum rules, such as the Bjorken \cite{Bj66},
Gerasimov-Drell-Hearn \cite{GDH66} or Burkhardt-Cottingham \cite{BC70}
sum rules, can then provide important tests of QCD and its applications
to nucleon structure.
Sum rules involving moments of neutron structure functions
(for example, the Bjorken sum rule, which relates the isovector
combination $g_1^p-g_1^n$ to the axial charge, $g_A$) require the
nuclear corrections to be known to a sufficient level of accuracy.

\begin{figure}[t]
\includegraphics[width=8cm]{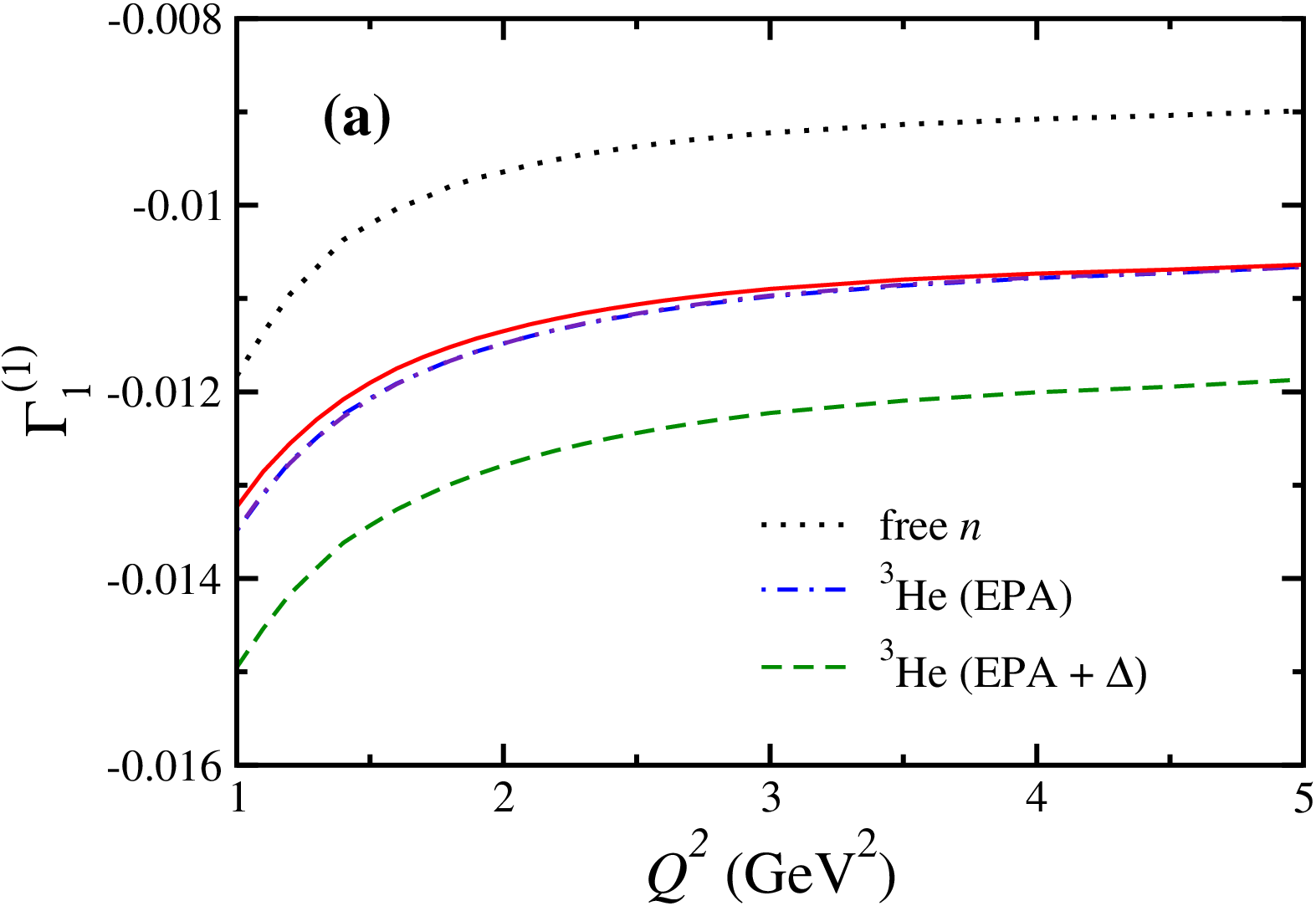}
\includegraphics[width=7.7cm]{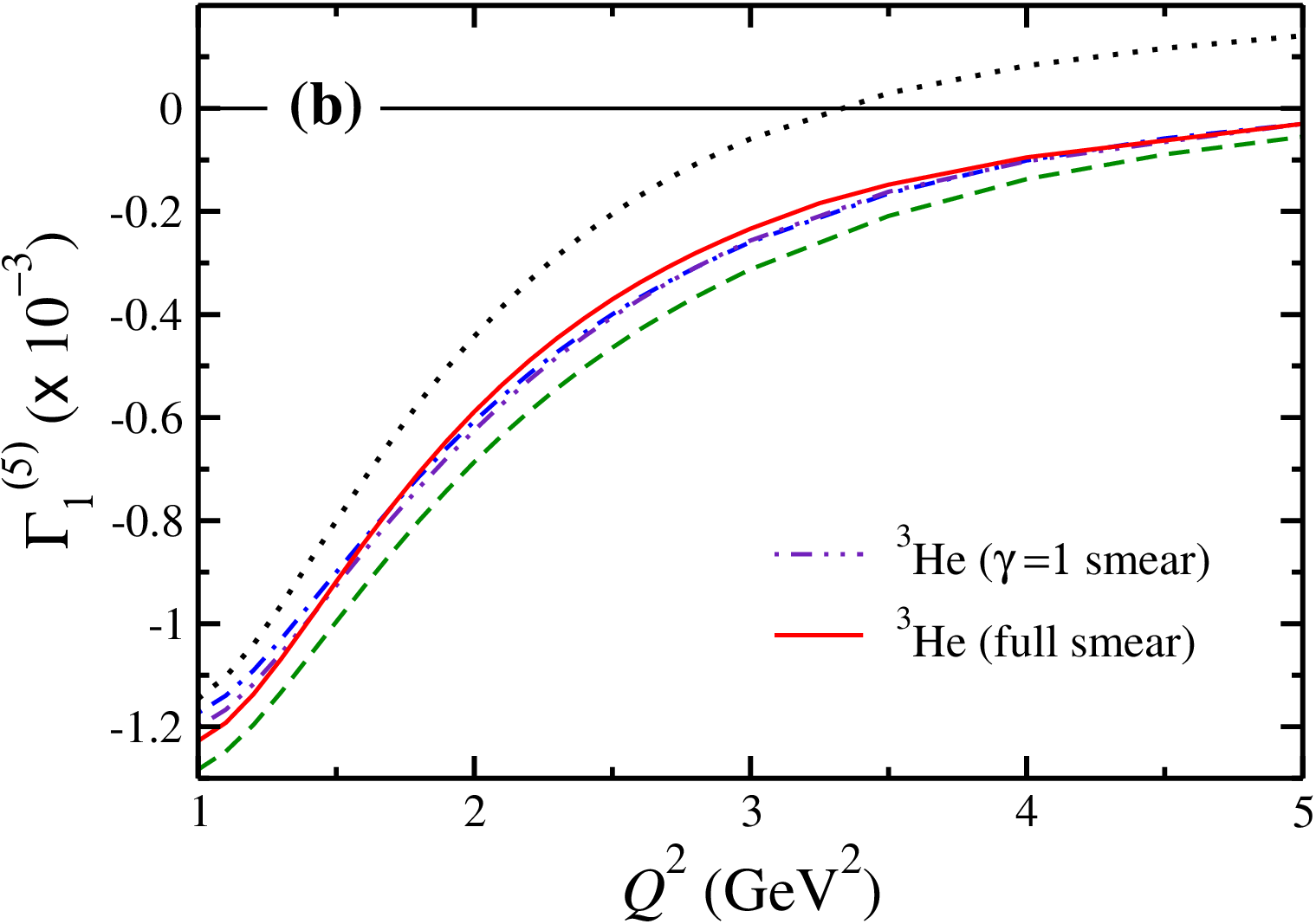}\\
\vspace*{0.9cm}
\includegraphics[width=8cm]{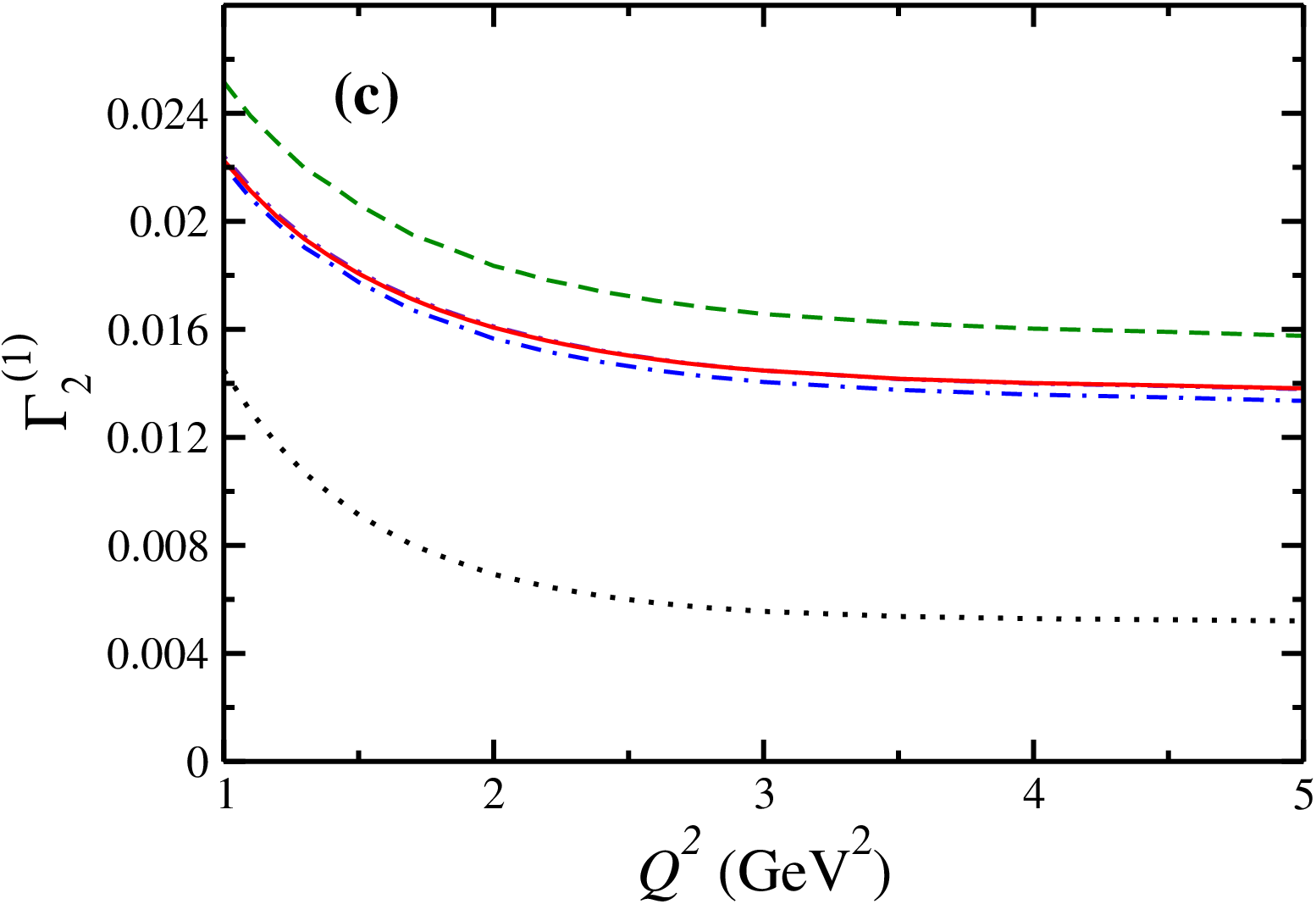}
\includegraphics[width=7.7cm]{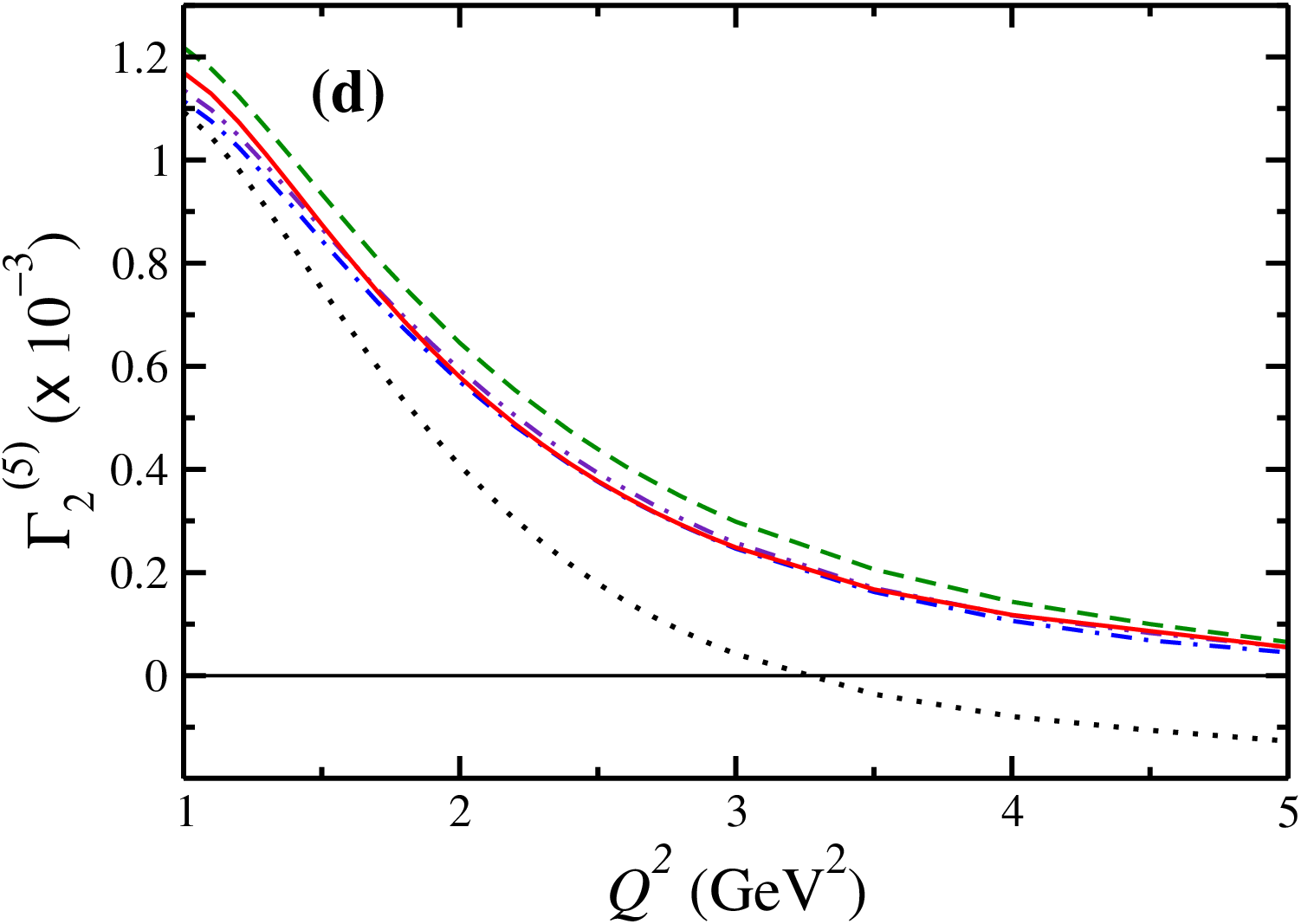}
\caption{Moments of the neutron and $^3$He $g_1$ structure
	functions, $\Gamma_1^{(n)}$ [{\bf (a)} and {\bf (b)}]
	and of the corresponding $g_2$ structure functions,
	$\Gamma_2^{(n)}$ [{\bf (c)} and {\bf (d)}],
	for $n=1$ and $n=5$.
	The $^3$He moments are computed in the EPA with nucleons only
	(dot-dashed) and with $\Delta$ components (dashed), and with
	Fermi smearing for $\gamma=1$ (dot-dot-dashed) and at finite
	$Q^2$ (solid).
	The $g_1$ moments are computed from the DSSV \cite{DSSV09}
	and MAID \cite{MAID99} parametrizations of the proton and
	neutron structure functions in the DIS and resonance regions,
	respectively, while the $g_2$ moments assume the WW relation
	(\ref{eq:WW}) for the DIS region and the MAID fit for the
	resonance part.}
\label{fig8}
\end{figure}

The effect of the nuclear corrections on the neutron $\Gamma_1^{(n)}$
and $\Gamma_2^{(n)}$ moments are illustrated in Fig.~\ref{fig8} for
the $n=1$ and $n=5$ moments from $Q^2 = 1$ to 5~GeV$^2$.
For the $g_1$ moments the DSSV \cite{DSSV09} and MAID \cite{MAID99}
parametrizations are used for the proton and neutron structure
functions in the DIS and resonance regions, respectively, while the
$g_2$ moments assume the WW relation (\ref{eq:WW}) for the DIS region
and the MAID fit for the resonance component.
For the lowest, $n=1$ moment computed within the EPA with nucleon
contributions only, the neutron effective polarization $P_1^n$ reduces
the magnitude of the (negative) neutron moment by $\sim 15\%$.
However, while the total proton polarization is small,
$2 P_1^p \sim -5\%$, the much larger value of the (positive) proton
moment $\Gamma_1^{p (1)}$ more than compensates, rendering the overall
correction to the $^3$He moment negative ($\sim 20\%$ larger magnitude).
Because the lowest moment is dominated by the small-$x$ contributions,
the effects of nuclear smearing are negligible, with only small
differences visible between the full, $Q^2$-dependent smearing
and that in the Bjorken ($\gamma=1$) limit.
More important is the contribution from the $\Delta$ resonance,
which is assumed in the EPA calculation of Sec.~\ref{ssec:Delta}
to be present at all $x$.  This gives a negative contribution to
the $^3$He moment which is comparable in magnitude to that from
the effective nucleon polarization correction.

Small-$x$ contributions are suppressed for higher moments,
as seen in Fig.~\ref{fig8}(b) for the $n=5$ moment of $g_1$.
In this case the relative effect of the nuclear smearing is
enhanced, although not significantly, while the effect of the
$\Delta$ resonance correction is reduced compared with the
lowest moment.  Note that because of the suppression of the
small-$x$ region by the factor $x^4$ in Eq.~(\ref{eq:moment}),
the magnitude of the $\Gamma_1^{(5)}$ moment is smaller by
at least an order of magnitude compared with $\Gamma_1^{(1)}$.

The behavior of the $g_2$ moments in Figs.~\ref{fig8}(c) and (d)
is qualitatively similar to the $g_1$ moments.  Generally the
sign of the $g_2$ structure function and its moments are opposite
from that of $g_1$, but the overall effects of the various
approximations for the nuclear corrections are analogous.
Namely, the EPA raises the magnitude of $\Gamma_2^{(n)}$ for
$^3$He from the neutron value due to the overall positive proton
contribution (since the proton $\Gamma_2^{(n)}$ is negative),
with the $\Delta$ resonance contribution giving an additional
small increase.  The effect of the latter is reduced for the $n=5$
moment, and the effects of the smearing are again relatively small.
Note that the smearing effects preserve the vanishing of the lowest
($n=1$) moment of $g_2$,
  $\Gamma_2^{^3{\rm He}(1)}\Big|_{\rm WW} = 0$,
so that the nonzero values of $\Gamma_2^{(1)}$ in Fig.~\ref{fig8}(c)
are entirely due to the resonance contributions, which need not
satisfy the WW relation.

\begin{figure}[t]
\includegraphics[width=8cm]{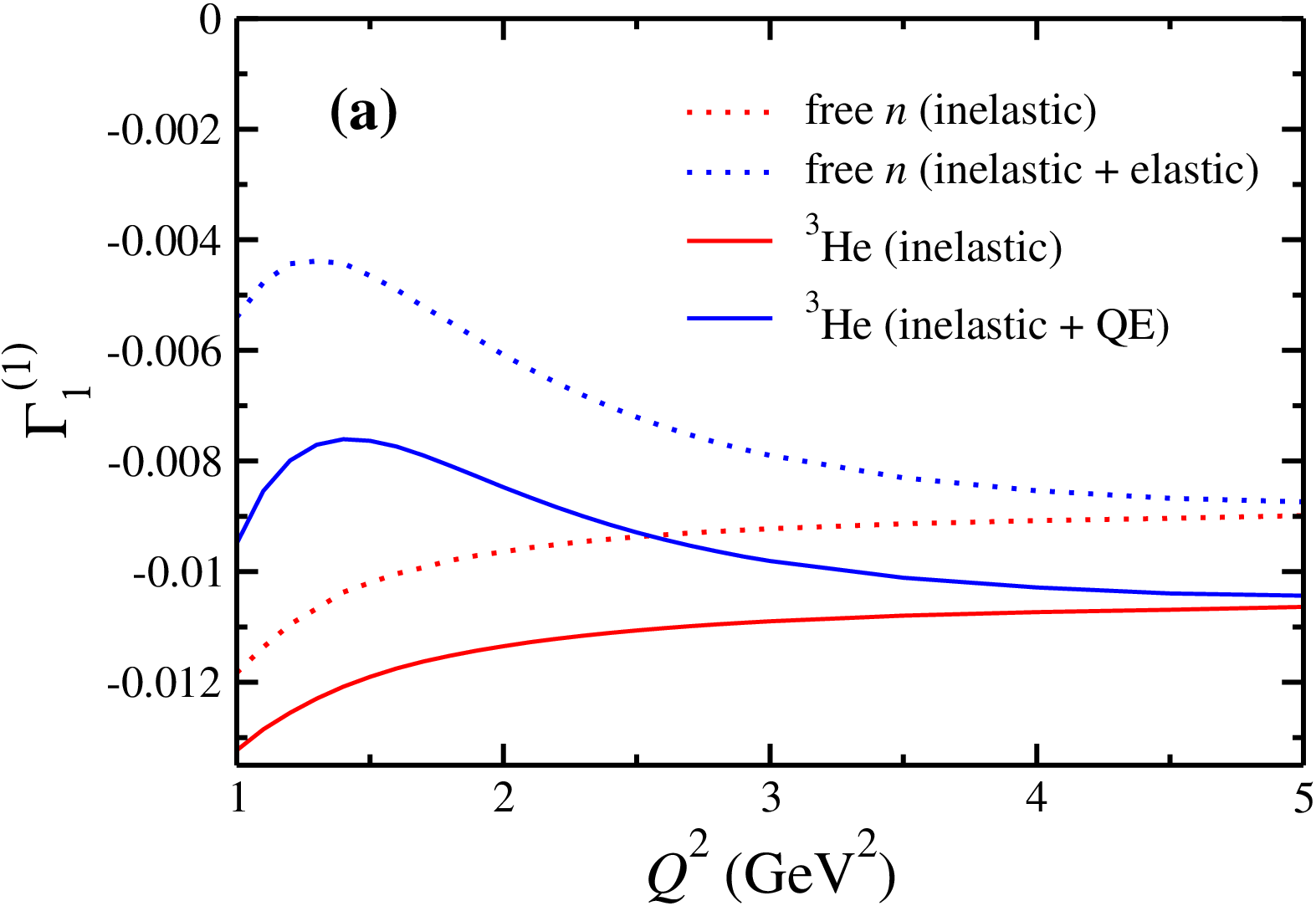}
\includegraphics[width=7.7cm]{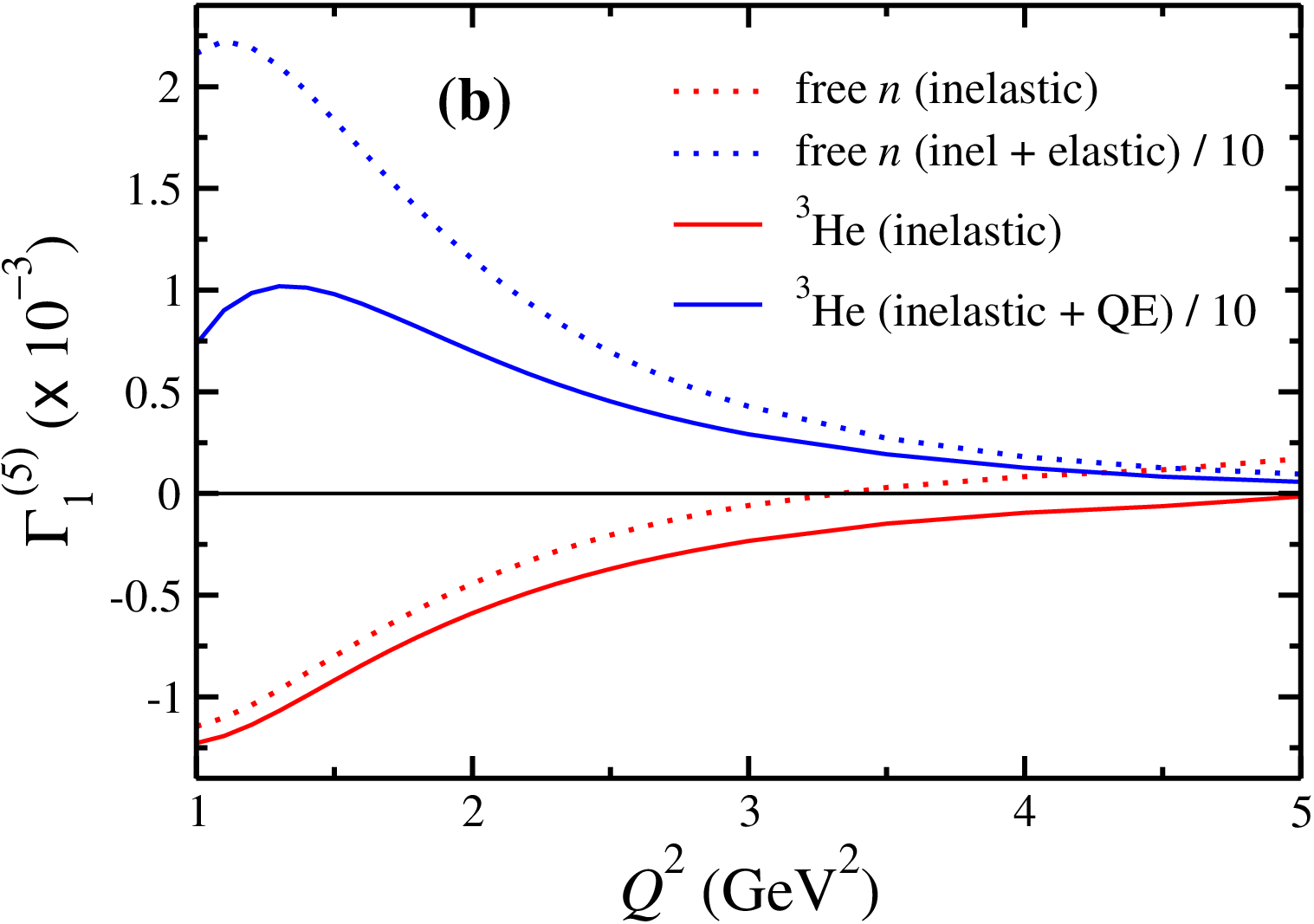}
\caption{Contributions from elastic and quasi-elastic scattering
	to the neutron and $^3$He $g_1$ moments, for
	{\bf (a)} $\Gamma_1^{(1)}$ and {\bf (b)} $\Gamma_1^{(5)}$,
	compared with the inelastic contributions.  The elastic and
	QE components are scaled by a factor 1/10 for clarity.}
% using combined MAID and DSSV parametrizations.}
\label{fig9}
\end{figure}

Since the moments of structure functions are formally defined as
integrals over the entire range of $x$ between 0 and 1, they in
principle contain contributions from elastic scattering at $x=1$
for the nucleon, and from QE scattering at $x \approx 1$ for
$^3$He.  The elastic and QE contributions are strongly suppressed
with increasing $Q^2$, but can be significant at
$Q^2 = {\cal O}$(1~GeV$^2$), as Fig.~\ref{fig9} illustrates
for the $\Gamma_1^{(1)}$ and $\Gamma_1^{(5)}$ moments.
As in Fig.~\ref{fig5}, the electromagnetic form factors of the
proton are taken from the parametrization of Ref.~\cite{AMT07},
and the neutron form factors from Ref.~\cite{Kelly04}, although
the dependence on the form factor fit is small.
For higher moments, the magnitude of the inelastic contributions
(at $x < 1$) is suppressed by the factor $x^n$, whereas the elastic
contribution (at $x = 1$) remains the same for all moments.
The QE contribution to the $\Gamma_1^{(5)}$ moment is therefore
significantly larger than the inelastic, especially at low $Q^2$
values, and for clarity in Fig.~\ref{fig9}(b) the sum of the
inelastic and elastic (or QE) is scaled by a factor 1/10.

\begin{figure}[t]
\includegraphics[width=8cm]{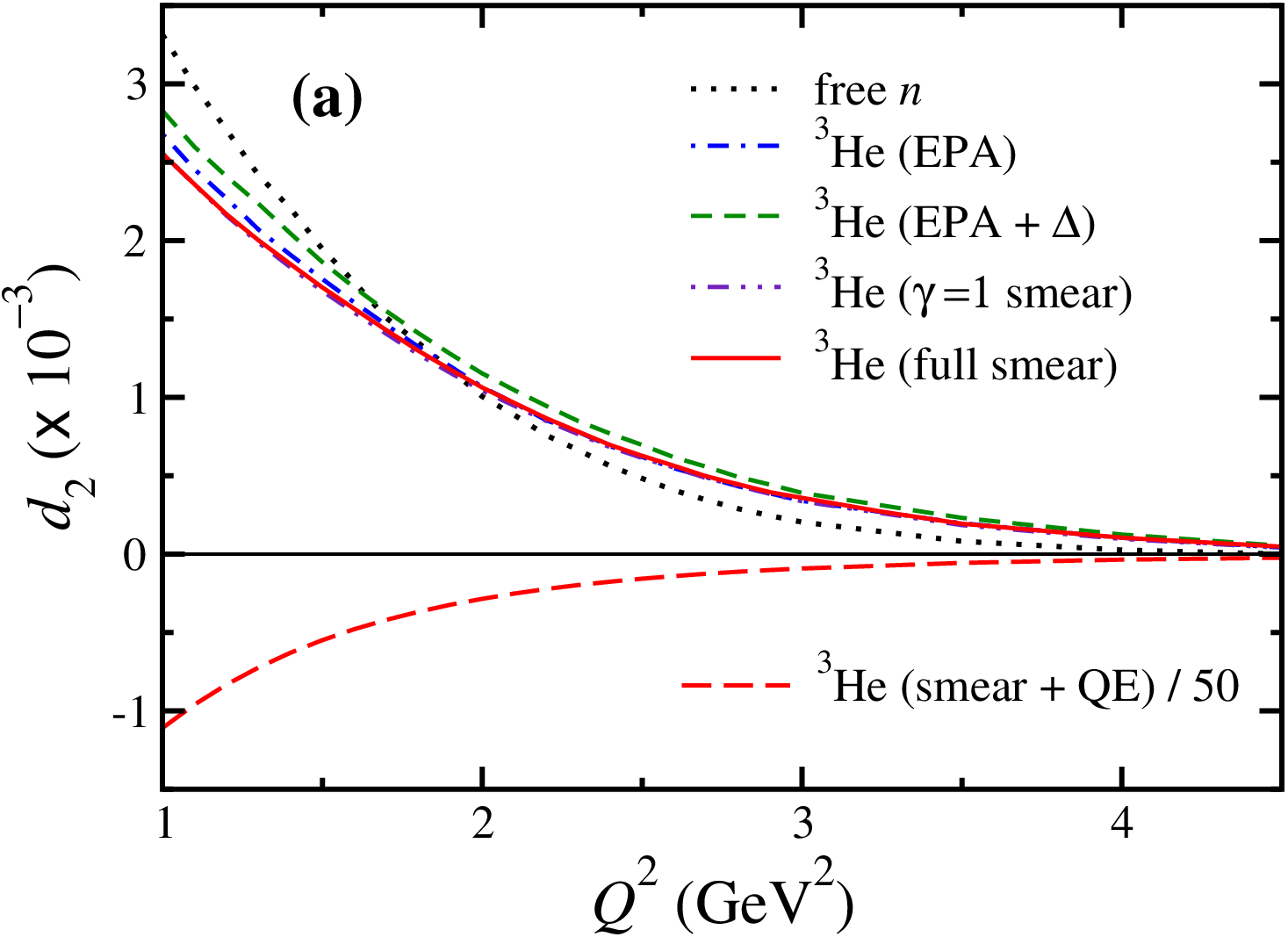}
\includegraphics[width=8cm]{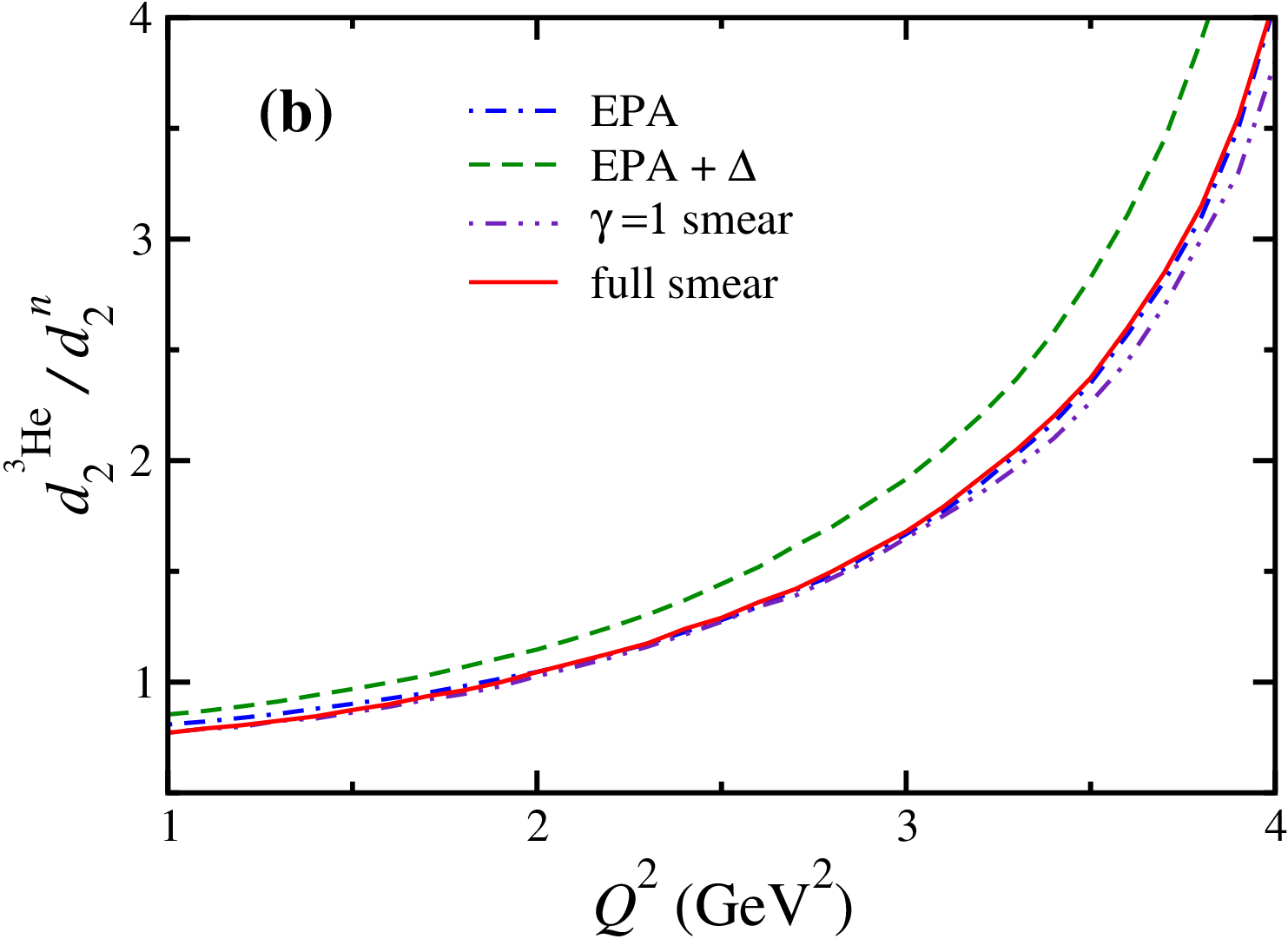}
\caption{{\bf (a)}
	$d_2$ moment of the neutron (dotted) and $^3$He, with the
	latter computed in the EPA with nucleons only (dot-dashed)
	and with $\Delta$ components (dashed), and with Fermi smearing
	for $\gamma=1$ (dot-dot-dashed) and the full smearing at
	finite $Q^2$ (solid).  The $d_2$ moment for $^3$He including
	the QE contribution (scaled by a factor 1/50) is shown for
	comparison (long-dashed).
	 {\bf (b)}
	Ratio of the $d_2$ moments for $^3$He and the neutron,
	with the $^3$He moments computed using the various
	approximations in (a).}
\label{fig10}
\end{figure}

Finally, to estimate the nuclear corrections to the $d_2$ moment
of the neutron defined in Eq.~(\ref{eq:d2}), in Fig.~\ref{fig10}
we show the $d_2$ moments for $^3$He computed using the various
approximations for the nuclear effects discussed above.
The $d_2$ moment is of interest because of its unique sensitivity
to higher twist contributions to the $g_2$ structure function
(the leading twist contribution from the WW relation vanishes,
as seen from Eqs.~(\ref{eq:d2}) and (\ref{eq:WW})).
The $d_2$ moment of the $^3$He structure functions was recently
measured at Jefferson Lab in the E06-014 experiment \cite{E06-014}.
The data are currently being analyzed, and are expected to have a
statistical precision of $\pm 0.4 \times 10^{-3}$ over the $Q^2$
range between 2 and 5~GeV$^2$, with an average
$\langle Q^2 \rangle \approx 4$~GeV$^2$ \cite{E06-014_pc}.

Using the MAID and DSSV parametrizations for $g_1$ and $g_2$
for the nucleon resonance and DIS regions, respectively, as in
Fig.~\ref{fig2} for instance, only the former makes a nonzero
contribution to the $d_2$ moment (the DSSV fit is performed
exclusively in terms of leading twist PDFs).
The MAID resonance fit gives rise to $d_2$ values which drop
precipitously with $Q^2$.  The nuclear corrections to $d_2$
are small in absolute terms, but increase dramatically with
$Q^2$ for the ratio $d_2^{^3{\rm He}}/d_2^n$ for all the models
considered, such that $d_2^{^3{\rm He}}$ is
$\approx 2$ times larger than $d_2^n$ at $Q^2 \approx 3$~GeV$^2$,
and
$\approx 4$ times larger at $Q^2 \approx 4$~GeV$^2$.
The effect of the nuclear smearing is minimal compared with
the EPA, although the possible $\Delta$ resonance component
of the $^3$He wave functions makes a non-negligible contribution
to the ratio in Fig.~\ref{fig10}(b).
Nucleon off-shell effects may also give rise to corrections to
$d_2^n$, however, these are difficult to estimate using the
(leading twist) quark models discussed in Sec.~\ref{ssec:off}.
At low $Q^2$, the QE contribution in Fig.~\ref{fig10}(a) is
significantly larger than the inelastic, and remains sizeable
at larger $Q^2$ also.  Accurate extraction of $d_2^n$ from the
$^3$He data will therefore require precise knowledge of the
nuclear effects and the elastic nucleon form factors over the
$Q^2$ range considered here.

%%%%%%%%%%%%%%%%%%%%%%%%%%%%%%%%%%%%%%%%%%%%%%%%%%%%%%%%%%%%%%%%%%%%%%%%%
\section{Conclusion}
\label{sec:conc}

For the foreseeable future, polarized $^3$He targets will remain
an essential tool for studying the spin structure of the nucleon,
providing the most direct means of probing the spin-dependent quark
and gluon distribution in the free neutron.  With the ever increasing
levels of precision attained in new generations of polarized DIS
experiments, including in previously unexplored regions of kinematics,
comes the need for correspondingly better understanding of the nuclear
effects that differentiate between the structure of the free neutron
and that bound in the $^3$He nucleus.

In this paper we have performed a comprehensive analysis of nuclear
corrections to the spin-dependent $g_1$ and $g_2$ structure functions
and their moments, as well as the $A_1$ and $A_2$ polarization
asymmetries which are also sensitive to nuclear effects in
unpolarized $^3$He structure functions.  
We have contrasted various methods of accounting for the nuclear
corrections, including through the use of effective polarizations,
and nuclear smearing functions computed in the framework of the
weak binding approximation.
Generally, the effective polarization approximation does not
provide a reliable means of describing the differences between the
$^3$He and neutron structure functions, especially in the low-$W$
region dominated by nucleon resonances and in the DIS region at
large values of $x$.  In these regions in particular it is important
to treat the $Q^2$ dependence in the smearing functions correctly,
as the comparison with the smearing computed in the Bjorken
($\gamma=1$) limit illustrates that the latter significantly
underestimates the strength of the effect.  On the other hand,
at intermediate $x$ values and at $W$ above the resonance region,
where the structure functions are smooth and slowly varying,
nuclear smearing provides only a relatively minor improvement
over the EPA approach.

In addition to the corrections arising from the incoherent nucleon
impulse approximation, we have also examined contributions from
non-nucleonic degrees of freedom in the nucleus, specifically the
$\Delta$ resonance.  Following Bissey {\it et al.} \cite{Bissey01},
we relate the strength of this correction to the Bjorken sum rule
in $A=3$ nuclei, and confirm sizeable contributions at small and
intermediate values of $x$, which consequently have greatest impact
on the lowest moments of the $g_1$ and $g_2$ structure functions.
Corrections associated with the nucleon off-shell structure have
also been estimated in a covariant spectator model, with the magnitude
determined by the change in the size of the nucleon radius in the
$^3$He nucleus, as well as from a quark-meson coupling model.
In both cases the off-shell corrections were found to cancel
somewhat the effects of the $\Delta$ contribution, although these
corrections at present are difficult to quantify model-independently.

Our analysis complements earlier studies of nuclear corrections to
spin-dependent structure functions, where some of these effects
were partially explored.  It also provides estimates of the nuclear
corrections to the $d_2$ moment of the neutron, measured recently
in the E06-014 experiment at Jefferson Lab \cite{E06-014}, which
offers a direct window on the higher twist component of the $g_2$
structure function.  The QE contribution to the $d_2$ moment of
$^3$He is found to be significant, requiring this component to be
determined to a high level of accuracy when extracting the neutron
$d_2$ results.

Measurement of the QE contributions to the polarized inclusive
$^3$He cross sections can in future provide an important test of the
nucleon smearing functions in $^3$He.  We have found non-trivial
cancellations between QE proton and neutron contributions to the
$g_1$ structure function of $^3$He, which is particularly striking
at intermediate values of $Q^2 \sim 1$~GeV$^2$.

While the goal of many $^3$He DIS experiments is ultimately the
extraction of information on the structure of the free neutron,
this is relatively straightforward only for moments of the structure
functions.  Our calculations of the nuclear corrections should
provide a reliable estimate of the size of these corrections and
their uncertainties.  Extraction of the neutron polarization
asymmetries $A_{1,2}^n$ and structure functions $g_{1,2}^n$ is
more challenging, on the other hand, especially in the nucleon
resonance region.  Here this will require unfolding the neutron
structure information by making use of a deconvolution procedure,
stepping through several iterations until convergence is achieved.
As found by Kahn {\it et al.} \cite{Kahn09}, typically this involves
just a handful of iterations, depending on the level of accuracy
required in the reconstruction, although precision data are needed to
obtain errors comparable to those for the free proton \cite{Malace10}.

Definitive tests of the nuclear correction methods would be possible
through independent determination of the free neutron structure in
experiments where the nuclear effects are minimal or absent altogether
\cite{TownHall}.
Examples of such processes include the polarized version of the
MARATHON proposal \cite{MARATHON} at Jefferson Lab, which will measure
the ratio of inclusive $^3$He to $^3$H structure functions, from which
the $d$ to $u$ quark PDF ratio will be extracted.  For unpolarized
scattering, nuclear corrections were found \cite{Afnan00} to cancel
to within $\approx 1\%$ up to $x \approx 0.85$, and similar effect
are expected for the spin-dependent case.
An alternative approach would be to perform semi-inclusive DIS from
polarized $^3$He, with detection of correlated $pp$ pairs that would
indicate scattering from the bound neutron.  Detection of such pairs
with low momentum at backward angles would minimize the degree to
which the struck neutron was off shell and eliminate contamination
from final state interactions, in analogy with the BONuS experiment
at Jefferson Lab with an unpolarized deuteron target \cite{BONuS}.
A more challenging method that would be completely free of nuclear
contamination would be parity-violating DIS of unpolarized leptons
from polarized protons \cite{Anselmino94, Hobbs08}.  The polarization
asymmetry here would be sensitive to the spin-dependent $\gamma Z$
interference structure functions, thus providing an independent
combination of the polarized $\Delta u$ and $\Delta d$ PDFs at
large $x$ from which the free neutron structure function could
be unambiguously reconstructed.

%%%%%%%%%%%%%%%%%%%%%%%%%%%%%%%%%%%%%%%%%%%%%%%%%%%%%%%%%%%%%%%%%%%%%%%%%
\begin{acknowledgments}

We thank S.~Kulagin for helpful discussions about nuclear effects in
$^3$He, L.~Brady for assistance with the nucleon off-shell corrections,
and D.~Parno for discussions about the E06-014 data.
This work was supported by the U.S. Department of Energy contract
No.~DE-AC05-06OR23177, under which Jefferson Science Associates, LLC
operates Jefferson Lab.  J.E. was partially supported by the SULI
program of the DOE, Office of Science.

\end{acknowledgments}

%%%%%%%%%%%%%%%%%%%%%%%%%%%%%%%%%%%%%%%%%%%%%%%%%%%%%%%%%%%%%%%%%%%%%%%%%

\end{document}